\documentclass[useAMS,usenatbib,letterpaper]{mn2e}
\usepackage[totalwidth=480pt,totalheight=680pt]{geometry}
\usepackage{graphicx,amssymb,amsbsy}
\input psfig

\title[The Solar System's Post-Main Sequence Escape Boundary]
{The Solar System's Post-Main Sequence Escape Boundary}
\author[Veras \& Wyatt]{Dimitri Veras$^{1}$\thanks{E-mail:veras@ast.cam.ac.uk}, Mark C. Wyatt$^{1}$\\
$^{1}$Institute of Astronomy, University of Cambridge, Madingley Road, Cambridge CB3 0HA}

\begin{document}

\date{Accepted 2012 January 9.  Received 2012 January 4; in original form 2011 December 2}

\pagerange{\pageref{firstpage}--\pageref{lastpage}} \pubyear{2012} 

\maketitle

\label{firstpage}

\begin{abstract}
The Sun will eventually lose about half of its current mass nonlinearly
over several phases of post-main sequence evolution.  This mass loss will
cause any surviving orbiting body to increase its semimajor axis and perhaps vary its  
eccentricity.  Here, we use a range of Solar models spanning plausible evolutionary
sequences and assume isotropic mass loss to assess the possibility of escape from the
Solar System.  We find that the critical semimajor axis in the Solar System 
within which an orbiting body is guaranteed to remain bound to the dying Sun
due to perturbations from stellar mass loss alone is $\approx 10^3$ AU - $10^4$ AU.
The fate of objects near or beyond this critical semimajor
axis, such as the Oort Cloud, outer scattered disc and specific bodies such as Sedna,
will significantly depend on their locations along their orbits 
when the Sun turns off of the main sequence. These results are applicable to any
exoplanetary system containing a single star with a mass, metallicity and age which
are approximately equal to the Sun's, and suggest that few extrasolar Oort Clouds
could survive post-main sequence evolution intact.
\end{abstract}

\begin{keywords}
planet-star interactions, planets and satellites: dynamical evolution and stability, stars: evolution, stars: AGB and post-AGB, Oort Cloud, minor planets, asteroids:general
\end{keywords}

\section{Introduction}

The fate of our Solar System is of intrinsic human interest.
From the present day, the Sun will continue to exist in a main 
sequence phase for at least 6 Gyr longer, and then experience 
post-main sequence phases before ending its life as a 
white dwarf.  The post-main sequence phases will be violent: the 
Sun's radius and luminosity will likely vary 
by several orders of magnitude while the Sun ejects approximately half 
of its present-day mass through intense winds.  The effect on orbiting
bodies, including planets, asteroids, comets and dust, may be 
catastrophic.

The future evolution of the 
Solar System between now and the end of the
Sun's main sequence will be largely unaffected by the Solar mass loss.
The Sun's current mass loss rate lies in the range $\sim 10^{-14} M_{\odot}/$yr - $10^{-13} M_{\odot}/$yr 
\citep[e.g.][]{cohen2011,pitpit2011}.
As demonstrated by \cite{veretal2011}, all orbiting objects would evolve
adiabatically due to this small mass loss.  Thus, assuming the mass loss rate remains
constant by the end of the main sequence,
their eccentricities will remain unchanged, and their semimajor axes would 
increase by at most $\approx 0.055\%$.  The Sun's main sequence mass loss 
rate, has, however, varied over time.  \cite{woodetal2002} and \cite{zenetal2010} suggest that 
this mass loss rate was orders of magnitude greater when the Sun began life 
on the main sequence, and is a monotonically decreasing function of time.  
Therefore, $0.055\%$ represents an upper bound.  Regardless, this 
potentially slight orbital expansion, uniformly applied to all orbiting 
objects, is not predicted to change the dynamics of the Solar System.

Instead, the primary driver of dynamical change in the Solar System 
during the Sun's main sequence will 
arise from the mutual secular perturbations of the orbiting bodies.
\cite{khokuz2007} provide
a comprehensive review of the investigations up until the year 2007 
which have contributed to our understanding of this evolution.  These
studies describe orbital evolution beyond $10^4$ yrs from
now in a primarily {\it qualitative} manner because numerical integrations
typically cannot retain the orbital phase information of the terrestrial
planets over the Sun's main sequence lifetime.
Further, Mercury might suffer a close encounter with Venus \citep{laskar2008}, and the 
orbital motion of the other inner planets
is predicted to be chaotic.  The outer planets, however, are predicted to survive and harbor
quasi-periodic motion.  Since this review, additional studies have primarily 
focused on the complex evolution of the terrestrial planets; \cite{lasetal2011} helps
affirm the intractability of terrestrial orbital phases
over long times.  Other studies 
(\citealt*{batlau2008} and \citealt*{laskar2008})
corroborate previous results about how the outer Solar System planets 
(Jupiter, Saturn, Uranus, and Neptune) should remain stable in 
near-circular ($e < 0.1$) orbits, even though the orbits themselves may be 
chaotic \citep{hayes2007,hayes2008,hayesetal2010}.

This uncertainty of a planet's orbital architecture at 
the beginning of the Sun's post main sequence phases renders difficult the 
determination of the fate of individual presently-known bodies.  
Regardless, several studies have explored the effect of the Sun's 
post main-sequence evolution on the planets.
The outcome for the Earth is controversial because investigators differ 
on whether the Earth's expanding orbit will ``outrun''
the Sun's expanding envelope, and on how to model the resulting tidal 
interactions (see \citealt*{schcon2008} and references therein).
\cite{dunlis1998} consider 
the stability and orbits of Jupiter, Saturn, Uranus and Neptune
with mass scalings and a post main-sequence 
mass loss prescription from \cite{sacetal1993}.  They find that
this prescription yields ``no large growth in planetary eccentricities'',
but mention the possibility of Oort Cloud comets being 
ejected during periods of brief, rapid mass loss.  Complimentary studies
of other planetary systems have explored in more detail how
Oort Cloud comet-analogues may become unbound \citep{alcetal1986,paralc1998}.

Our study is devoted to exploring the prospects for dynamical ejection
from the Solar System by using realistic multi-phasic
nonlinear mass loss prescriptions from the SSE code \citep{huretal2000} and
by treating isolated objects beyond the influence of any potentially 
surviving planets.  We do so by using the analytical mass loss techniques 
in \cite{veretal2011}, which yield a critical 
semimajor axis, $a_{\rm crit}$, within which an orbiting object is guaranteed to remain bound.
In Section 2 we present the solar evolutionary models used, highlighting
those which are later applied to N-body simulations that help reaffirm the validity of 
$a_{\rm crit}$.  In Section 3, we define $a_{\rm crit}$, plot it as a function of Solar
evolution model, and explore the properties of some representative orbits.
In Section 4, we explore the implications for the scattered disc and the Oort Cloud.
We discuss the results in Section 5 and conclude in Section 6.

\begin{figure}
\centerline{
\psfig{figure=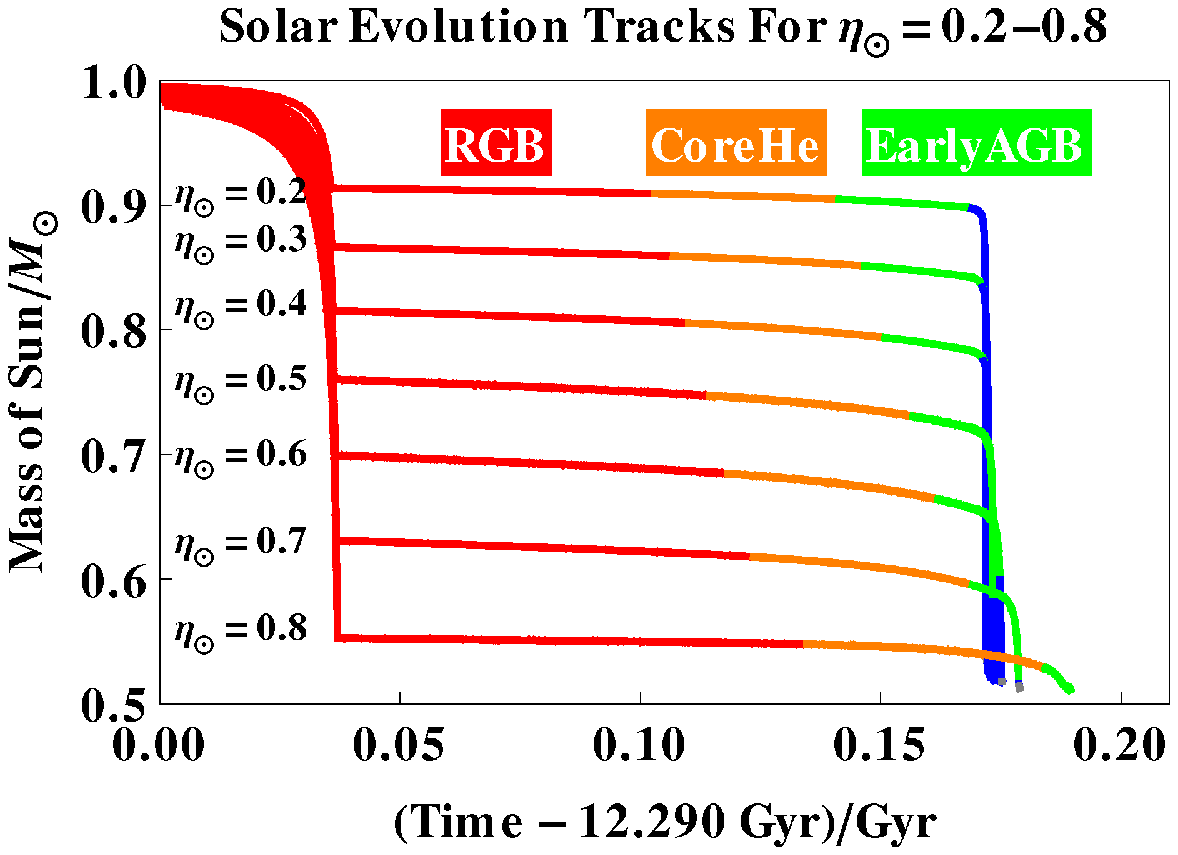,width=8.5cm}
}
\centerline{ }
\centerline{
\psfig{figure=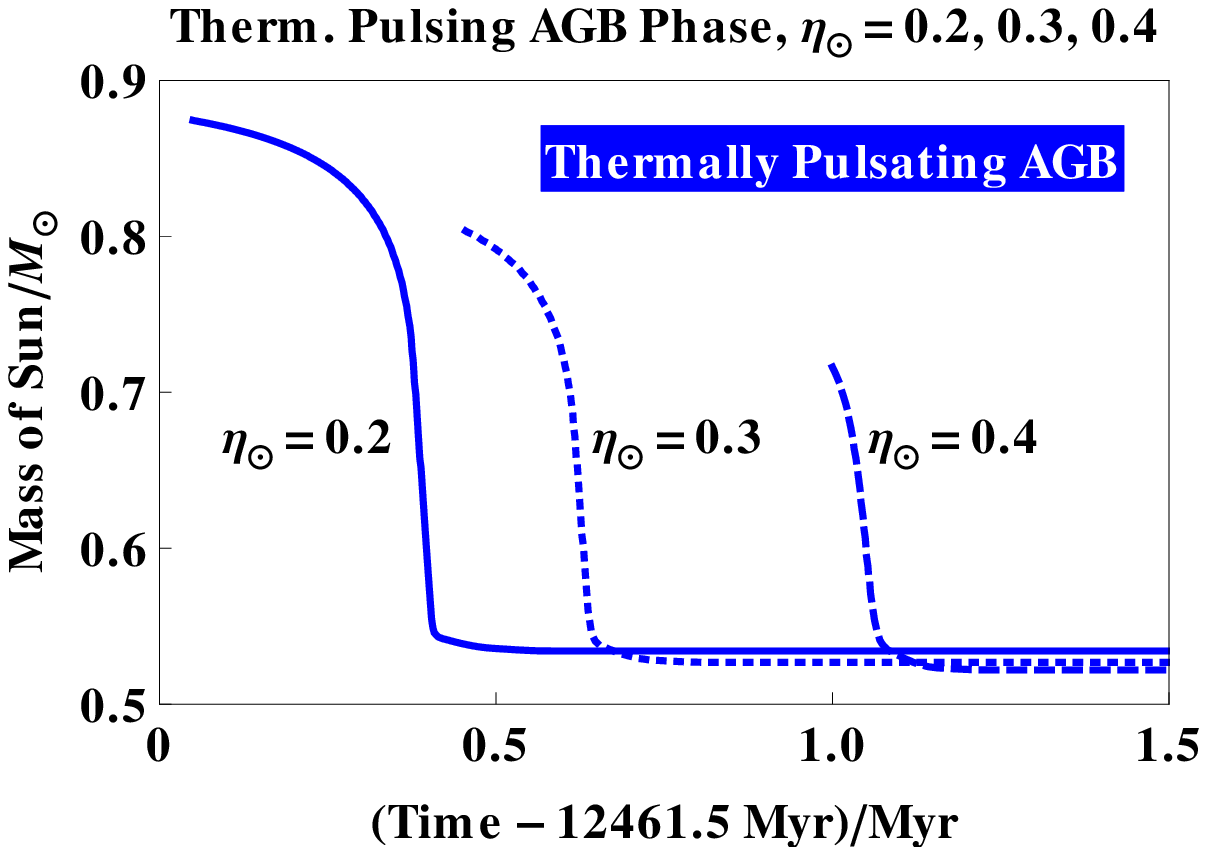,width=8.5cm} 
}
\caption{
Potential Solar evolutionary tracks.  The upper panel displays 
at least a part of all phases from the RGB phase to the thermally
pulsing AGB phase for 7 representative values of $\eta_{\odot}$ (the nearly-flat
RGB curves before 12.290 Gyr are not shown).  The lower
panel zooms-in on and isolates the thermally pulsing AGB phase for
$\eta_{\odot}=0.2$ (solid line), $\eta_{\odot}=0.3$ (dotted line)
and $\eta_{\odot}=0.4$ (dashed line). Time is measured
from the beginning of the Sun's main sequence.  The plot demonstrates how the choice
of $\eta_{\odot}$ can drastically alter the percentage of Solar mass lost and Solar
mass loss rate per phase.
}
\label{plot1}
\end{figure}

\section{Solar Evolution Models}

We utilize the SSE code \citep{huretal2000}, which yields a complete stellar evolutionary track
of a star for a given initial stellar mass, metallicity and mass loss prescription.
We denote the Sun's mass as $M_{\star}(t)$, and define a Solar Mass $\equiv 1 M_{\odot}$ to be the 
{\it current value} of the
Sun's mass.  We assume that the Sun's current metallicity is [Fe/H]$_{\odot} \equiv 0.02$.
We also assume that the Sun has already been evolving on the main sequence for exactly $4.6$ Gyr.  

The remaining, and unknown, constraint to include is the Sun's post-main sequence mass
loss.  On the Asymptotic Giant Branch (AGB),
we use the mass-loss prescription from \cite{vaswoo1993}:

\begin{equation}
\log{\frac{dM_{\star}}{dt}} = -11.4 + 0.0125 
\left[P - 100 {\rm max}\left(M_{\star} - 2.5, 0.0 \right)  \right] 
\end{equation}

\noindent{such} that $dM_{\star}/dt$ is computed in $M_{\odot}/$yr and where

\begin{equation}
\log{P} \equiv {\rm min}\left(3.3, -2.07 - 0.9 \log{M_{\star}} + 1.94 \log{R_{\star}} \right)
,
\end{equation}

\noindent{where} $P$ is computed in years and $R_{\star}$ is the radius of the Sun.

On the Red Giant Branch (RGB), mass loss is often modeled to have the 
following ``Reimers Law'' functional form \citep{kudrei1978}: 

\begin{equation}
\frac{dM_{\star}}{dt} = \eta \left(4 \times10^{-13}\right)
\frac{L_{\star}(t) R_{\star}(t)}{M_{\star}} \frac{M_{\odot}}{\rm yr}
\end{equation}

\noindent{where} $L_{\star}$ represents 
the Solar luminosity and $\eta$ is
an important dimensionless coefficient whose magnitude helps
determine the sequence of stellar evolutionary phases.  
\cite{sacetal1993} compute evolutionary models for the Sun,
and state that $0.4 < \eta < 0.8$ represent 
``reasonable mass-loss rates''.  This conforms with the
typically-used value of $\eta = 0.5$ 
\citep[e.g.][]{kudrei1978,huretal2000,bonwya2010}.
An updated version of the Reimers Law \citep{schcun2005}
contains two new multiplicative factors:

\begin{equation}
\left(\frac{T_{\rm eff}}{4000 {\rm K}} \right)^{3.5}
\left[1 + \frac{g_{\odot}}{4300 g_{\star}}\right]
\end{equation}

\noindent{where} $g_{\star}$ and $T_{\rm eff}$ are the star's
surface gravity and effective temperature, and $g_{\odot}$
is the current value of the Solar surface gravity.
The product of both these additional factors is not far from unity.
According to \cite{schcun2005}, who used the RGBs and Horizontal Branches
of 2 globular clusters with very different metallicities for calibration,
their effective value of $\eta$ in the original Reimers law would be
about $0.5$.

All these considerations dissuade us from selecting a single
value of $\eta_{\odot}$ and instead lead us to consider the 
entire range $0.2 \le \eta_{\odot} \le 0.8$; we sample $\eta_{\odot}$ values
in this range in increments of $0.01$.  Representative 
evolutionary sequences in this range are displayed in the upper panel
of Fig. \ref{plot1}.
In all cases, the Sun undergoes a $\gtrsim 0.5$ Gyr-long transition
towards the foot of the RGB after leaving the main sequence.
The Sun moves on to the RGB before enduring a period
of core helium burning and subsequent evolution on the early AGB, as defined by \cite{huretal2000}.  
Then, for  $\eta_{\odot} \le 0.74$,
the Sun will enter a thermally pulsing AGB phase before becoming
a white dwarf.  This intermediate phase is bypassed for $0.75 \le \eta_{\odot} \le 0.80$.

The different evolutionary sequences demonstrate that the greatest amount of mass lost
and the greatest mass loss rate may occur during the RGB, Early AGB, {\it or} Thermally
Pulsing AGB phases.  In all cases the total percent of the Sun's current mass
which is lost is $\approx 46.5\%$-$49.0\%$.  The value of $\eta$ determines how much
{\it envelope} mass is lost on the RGB, and hence, the amount of envelope mass
which remains at the start of the AGB.  For $\eta_{\odot} \lesssim 0.5$, most of this disposable mass is
reserved for the AGB, whereas for $\eta_{\odot} \gtrsim 0.5$, most of this mass is lost during the
RGB phase.  When $\eta$ is large enough ($\eta_{\odot} \ge 0.75$), then enough of the envelope
mass is lost during the RGB to cause the Sun to bypass the Thermally Pulsing AGB phase.
In the extreme case of $\eta_{\odot} \ge 0.94$ (not modeled here), the Sun would bypass Helium
burning and the AGB phases entirely, and would transition to a pure Helium white dwarf
directly from the RGB phase.

The lower panel zooms in and isolates the thermally pulsing AGB phase for
$\eta_{\odot}=0.2-0.4$, where the mass loss rate is greatest.  Notice that the timescale
for this mass loss is $\approx 0.1-0.3$ Myr, orders of magnitude shorter than the mass
loss on the RGB.

\section{The Critical Semimajor Axis}

\subsection{Equations}

A star which is expunging mass beyond an orbiting object 
will cause the latter's orbit to expand.   If the mass
loss is large enough during one orbital period, then deformation of the orbit
can cause the object to escape the system.  We quantify
these claims by defining a dimensionless 
``mass loss index'' \citep{veretal2011}:

\begin{eqnarray}
\Psi &\equiv& \frac{\rm mass \ loss \ timescale}{\rm orbital \ timescale}
= \frac{\alpha}{n\mu}
\nonumber
\\
&=& \frac{1}{2\pi} 
\left( \frac{\alpha}{1 M_{\odot}/{\rm yr}}\right)
\left( \frac{a}{1 {\rm AU}}\right)^{\frac{3}{2}}
\left( \frac{\mu}{1 M_{\odot}}\right)^{-\frac{3}{2}}
,
\label{mlindex}
\end{eqnarray}

\noindent{where} $\alpha$ represents the Solar mass loss rate, 
$a$ and $n$ represent the orbiting object's semimajor
axis and mean motion, and $\mu = M_{\star} + M_s$, where $M_s$ represents
the mass of the object.  We can assume $\mu \approx M_{\star}$ because 
every known object which orbits the Sun has a mass
that is $< 0.1\%$ of $1 M_{\odot}$.  However, observations cannot yet
exclude the possibility of the existence of a Jovian planet 
orbiting at a distance of at least several thousand AU \citep{iorio2011}.  
Therefore, ``undiscovered'' planets beyond $\sim 10^4$ AU might be massive
enough to make a nonnegligible contribution to $\mu$. 
Even if such a planet formed
within the orbit of Neptune, gravitational scattering could have placed
the planet in its present location \citep{veretal2009}.  Alternatively,
the planet could have previously been free-floating and subsequently captured by the Solar System.

Dynamical evolution critically depends on $\Psi$.
When $\Psi \ll 1$, the orbiting object evolves
``adiabatically'' such that $a$ increases but its 
eccentricity, $e$, remains constant.  Examples
of objects with $\Psi \ll 1$ include all 8 of the 
inner and outer planets, both currently (see Section 1)
and during any plausible post-main sequence Solar evolution
phase.  Alternatively, when $\Psi \gg 1$,
in the ``runaway'' regime,
$a$ continues to increase but now $e$ may achieve any value from zero to unity.
\cite{veretal2011} demonstrate that the bifurcation
point beyond which the motion is no longer adiabatic is not sharp and 
occurs at $\Psi_{\rm bif} \sim 0.1-1.0$.  In fact, if $\Psi$ is
instead defined as $\alpha T/\mu$, where $T$ is the orbital period of
the orbiting object, then this definition differs from Eq. (\ref{mlindex}) 
by a factor of $1/(2\pi) \approx 0.16$.  Therefore, both definitions
effectively bound the approximate location of the bifurcation point when $\Psi = 1$, and we 
consider both possibilities in our analysis.

The Sun loses mass nonlinearly, as showcased in Fig. \ref{plot1}.
One can apply the analytical linear results
from \cite{veretal2011} to a nonlinear mass loss profile by partitioning the profile
into $N$ smaller approximately linear stages, with mass loss rates 
of $\alpha_1, \alpha_2, ...\alpha_i... \alpha_N$.  
If all these segments occur in the adiabatic regime, then we can relate the
semimajor axis and eccentricity at the start of the first segment to these same elements at
the end of segment $i$ with:

\begin{equation}
a_i = a_0 \left( \frac{\mu_i}{\mu_0} \right)^{-1} \equiv a_0 \beta_{i}^{-1}
,
\label{aadia}
\end{equation}

\begin{equation}
e_i = e_0
,
\label{eadia}
\end{equation}

\noindent{Because} the orbiting object's eccentricity does not change in this regime,
the body will remain bound in the absence of additional perturbations.

All orbiting bodies will be expanding their semimajor axis at the
same rate.  The mass loss index at the end of segment $i$ will be:

\begin{equation}
\Psi_i = \kappa \alpha_i \left( \frac{a_0 \mu_0}{\mu_{i}^2} \right)^{\frac{3}{2}}
\label{psii}
\end{equation}

\noindent{where} $\kappa = 1$ or $2\pi$, depending on whether
$\Psi$ is defined with respect to the mean 
motion (Eq. \ref{mlindex}) or the orbital period.
$\kappa$ could also be treated generally as a tunable
parameter which may better pinpoint the location of the bifurcation
point in specific cases.  The {\it first} instance
at which the orbiting object {\it may} escape is when the
evolution becomes non-adiabatic and hence the eccentricity
can vary.  This occurs when $\Psi_i \approx 1$.
Therefore, the minimum percent of $\mu_0$ which
must be retained in order to guarantee that the orbiting object
will remain bound in a given segment $i$ is:

\begin{eqnarray}
\beta_{{\rm crit}_i} &\equiv& \left( \frac{\mu_i}{\mu_0} \right)_{\rm crit}
\nonumber
\\
&\approx&
\kappa^{\frac{1}{3}}
\left( \frac{\alpha_i}{1 M_{\odot}/{\rm yr}}\right)^{\frac{1}{3}}
\left( \frac{a_0}{1 {\rm AU}}\right)^{\frac{1}{2}}
\left( \frac{\mu_0}{1 M_{\odot}}\right)^{-\frac{1}{2}}
.
\label{crit}
\end{eqnarray}

\noindent{Hence}, the critical $a_0$, termed $a_{\rm crit}$, at which an orbiting object
will remain in the adiabatic regime, and thus bound to a 1 $M_{\odot}$ star,
is

\begin{equation}
\frac{a_{\rm crit}}{1 {\rm AU}} 
\approx
\kappa^{-\frac{2}{3}}
{\rm min}
\left[
\beta_{i}^2
\left( \frac{\alpha_i}{1 M_{\odot}/{\rm yr}}\right)^{-\frac{2}{3}}
\right]
,
\label{acrit}
\end{equation}

\noindent{where} the minimum is taken over all segments $i$.

\subsection{Mass Loss Profile Partition}

The accuracy of $a_{\rm crit}$ depends on how finely the Solar mass
loss profile is partitioned into stages.  
Given the focused nature of our study,
we simply adopt the most accurate option by treating
each SSE timestep as a segment.  These segments 
also include transitions between evolutionary phases, for which 
a mass loss and timescale is computed by SSE in every case.
We choose output parameters which yield 3333 outputs total
per simulation for the RGB, core helium burning, 
early AGB and thermally pulsing AGB phases,
and at least 20 outputs for each of the other phases of evolution.  
Our choices are motivated by the observation that the 
segment which satisfies Eq. (\ref{acrit}) 
always arises from the RGB or an AGB phase.
The minimum SSE timesteps are $\approx 10^3$ yr long, which
is shorter than the orbital period of the objects we consider here.
Equation (\ref{acrit}) is valid for any relation
between SSE timesteps and orbital period because, in the adiabatic regime,
the semimajor axis and eccentricity evolution is independent of the location
of the object along its orbit.
We apply this output parametrization
uniformly across the $\eta_{\odot}$ values sampled.

\subsection{N-body Simulations} \label{Nbodsec}

We supplement the computation of $a_{\rm crit}$ (from the
stellar evolution profiles alone) with N-body simulations
of orbiting objects.  These simulations help affirm that objects with $a < a_{\rm crit}$
do remain bound, and provide a rough characterization of the
motion of objects which are susceptible to escape.  We 
select the 7 Solar evolutionary pathways corresponding to 
$\eta_{\odot} = [0.2, 0.3, 0.4, 0.5, 0.6, 0.7, 0.8]$ for these simulations,
and use a second-order mixed-variable symplectic integrator 
from a modified version of the {\it Mercury} integration
package \citep{chambers1999}.  All orbiting objects are treated
as test particles for computational ease and so that they
could be integrated simultaneously.  Because $\mu$ is dominated
by $M_{\star}$, an isolated giant planet
would react to stellar mass loss nearly equivalently to an isolated
test particle on the same orbit.  After each {\it Mercury} timestep,
we linearly interpolate the value of the Sun's mass from the SSE output.
This approximation is sufficient for numerical integrations which
don't feature close encounters.

Simulating orbital evolution throughout the remaining lifetime of the Solar 
System with numerical integrations and without utilizing secular
approximations is computationally unfeasible and, 
for this project, unnecessary.  Our focus is on the behavior
of objects when they are most susceptible to escape.  Therefore,
having already identified when the greatest mass loss occurs,
we simulate the entire thermally pulsating AGB phase for
the $\eta_{\odot} = [0.2, 0.3, 0.4]$ simulations (see Fig. \ref{plot1}).
The duration of these simulations is between 
$2.7 \times 10^5$ yr and $5.7 \times 10^5$ yr.
For $\eta_{\odot} = [0.5, 0.6, 0.7, 0.8]$, we simulate 
Solar mass evolution from when $M_{\star} = 0.98000 M_{\odot}$,
which occurs during the RGB, until 
the beginning of the white dwarf phase (see Fig. \ref{plot1}).
For these simulations, modeling the phases after
the RGB may be important because 
despite their overall low mass loss rates,
the AGB phases could feature short bursts of high mass loss.  
Also, $\beta_i$ is lower
for the phases beyond the RGB.
The duration of these simulations is between
$1.64 \times 10^8$ yr - $1.88 \times 10^8$ yr.

We choose the initial conditions for the orbiting objects as follows.
The orbital evolution of an orbiting object which is susceptible
to escape is highly dependent on $e$ and its true anomaly, $f$
(see \citealt*{veretal2011}).  For linear mass loss and
$\Psi \gg 1$, the object's eccentricity is likely 
to initially increase.  However, the eccentricity
must initially decrease if  
$f_{\rm crit} \le f \le \left(360^{\circ} - f_{\rm crit}\right)$, 
where $f_{\rm crit} = 180^{\circ} - (1/2)\cos^{-1}(7/9) \approx 160^{\circ}$,
and we wish to sufficiently sample this behavior as well.
These considerations motivate us to choose 
$f_0$ in increments of $30^{\circ}$ from $0^{\circ}-330^{\circ}$, supplemented
by $f_0 = [160^{\circ}, 170^{\circ}, 190^{\circ}, 200^{\circ}]$.
In order to sample a wide range of eccentricties, we choose
$e_0 = [0.0, 0.2, 0.4, 0.6, 0.8, 0.95]$.  Because application
of Eq. (\ref{acrit}) demonstrates that 
$10^3$ AU $\lesssim a_{\rm crit} \lesssim 10^4$ AU,
we choose $a_0/$AU$ = [10^2, 10^3, 3 \times 10^3, 5 \times 10^3, 10^4, 10^5]$.
The smallest of these values yields a period of 414 years 
for $M_{\star} = 0.5 M_{\odot}$.  Therefore, we choose a timestep
less than $1/50$th of this value, $1 \times 10^4$ days, for all 
our numerical simulations.  Because this timestep might
not sample the pericenter sufficiently for highly eccentric
orbits, we perform an identical set of simulations
with a timestep which is one order of magnitude lower
as a check on the results.  

\begin{figure*}
\centerline{
\psfig{figure=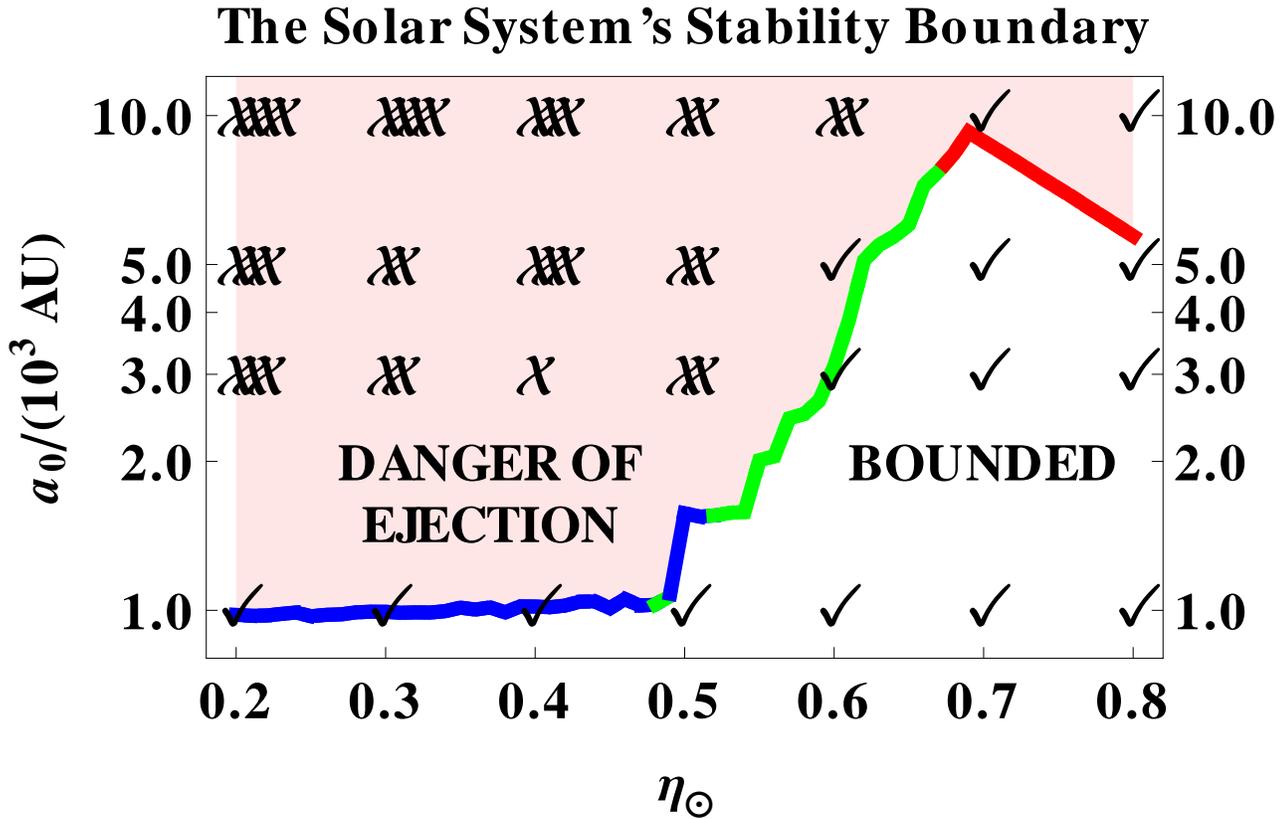,width=18cm} 
}
\caption{The Solar System's critical semimajor axis, as a function
of the Solar Reimers Mass Loss coefficient, $\eta_{\odot}$.  The curve is derived from
Eq. (\ref{acrit}).  In the shaded pink region
above the curve, orbiting bodies may escape the dying Sun on hyperbolic orbits.  In 
the white region below
the curve, bodies are guaranteed to remain bound on elliptical orbits. The colour of the line segments
indicates the Solar phase in which orbiting bodies are most susceptible to escape:
red = RGB, green = early AGB and blue = thermally pulsing AGB.  Checkmarks and crosses provide information about
the results of N-body integrations (see Section \ref{Nbodsec}).  Checkmarks indicate that every orbiting object in every
N-body simulation remained bound.  Each cross represents a different initial
eccentricity value (out of the set $\lbrace0.0, 0.2, 0.4, 0.6, 0.8, 0.95\rbrace$) 
of at least one body which escaped.  An alternate curve, with $\kappa = 2\pi$,
would have the same form but with values which are $(2\pi)^{2/3} \approx 3.4$ times
lower than the ones shown here, yielding a more conservative estimate.  
This plot demonstrates that Solar System bodies must satisfy $a < 10^3$ AU
when the Sun turns off of the main sequence in order to be guaranteed protection
from escape.
}
\label{plot3}
\end{figure*}

\subsection{Visualizing the Critical Semimajor Axis}
 
We plot $a_{\rm crit}$ from Eq. (\ref{acrit}) assuming $\kappa = 1$ as a function 
of $\eta_{\odot}$ in Fig. \ref{plot3}.  The curve illustrates that for $\eta_{\odot} \le 0.5$,
$a_{\rm crit} \approx 10^3$ AU, and for $\eta_{\odot} = 0.7$, $a_{\rm crit} \approx 10^4$ AU.
Other values of $a_{\rm crit}$ lie in-between these two extremes.  The curve is coloured
according to the stellar evolutionary phase at which an orbiting object is most susceptible to ejection,
which depends on a combination of the mass loss rate and the mass remaining in the Sun.
Three different phases (red = RGB, green = early AGB, blue = thermally pulsing AGB) are 
represented on the curve, demonstrating the sensitivity of $a_{\rm crit}$ to the stellar
model chosen.  When the RGB evolution dominates 
the orbiting object's motion, the critical semimajor axis is higher than in the AGB-dominated
cases because RGB evolution
takes place earlier in the Sun's post-main sequence life, when the Sun has more mass.
Any objects in the pink shaded region above the curve might escape; those in the white
region below the curve are guaranteed to remain bound.

The overlaid symbols indicate the results of the N-body simulations.  Checkmarks
indicate that every orbiting object from every $\lbrace e_0,f_0 \rbrace$ bin sampled 
remained bound ($e < 1$) for the duration of the simulation.  Crosses indicate
that at least one orbiting object become unbound, either through the orbit becoming
hyperbolic ($e > 1$) or through the body achieving 
$a \gtrsim 10^6$ AU.  The number of crosses indicate
the number of $e_0$ bins in which at least one orbiting object became unbound.  The N-body
simulations help affirm that all of the objects below the curve remained bound.
This also holds true if $\kappa = 2\pi$.  In that case, the curve would keep the same
form but be lowered by a factor of $(2\pi)^{2/3} \approx 3.4$.  This value of $\kappa$ would provide
a more conservative estimate for the Solar System's critical semimajor axis,
and would still ensure that any known surviving planets and belts would remain 
bound.  Objects
with high values of $e_0$ are usually, but not always, prone to escape;
additional cross symbols suggest that objects escaped at progressively 
lower values of $e_0$.  The number of crosses displayed do not show uniform
patterns primarily because of the complex dependencies of the evolution
in the adiabatic/runaway transitional regime on $a$, $e$ and $f$ (see Section 4.2).  Also,
the parameter space sampled is limited and the N-body simulations do not model
the entire pre-white dwarf lifetime of the Sun.

Not shown in Fig. \ref{plot3} are the results from the $a_0 = 10^2$ AU and
$a_0 = 10^5$ AU simulations.  In the former, every orbiting object remains bound
(corresponding to a row of checkmarks).
In the latter, at least one orbiting object escapes for each value of $\eta_{\odot}$
integrated (corresponding to a row of different numbers of crosses).  

\subsection{Properties of the Critical Stage}

\begin{figure}
\centerline{
\psfig{figure=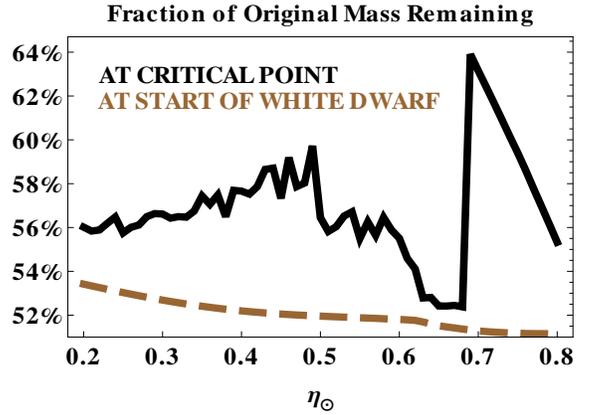,width=8.5cm,height=5.5cm} 
}
\caption{The percent of $1 M_{\odot}$ remaining in the Sun
at the moment when orbiting bodies are most likely to escape (solid black
curve) and at the moment the Sun becomes a white dwarf
(dashed brown curve).  Orbiting bodies are most likely to escape the 
Solar System after the Sun has lost at least $0.35 M_{\odot}$.
}
\label{plot4}
\end{figure}

\begin{figure}
\centerline{
\psfig{figure=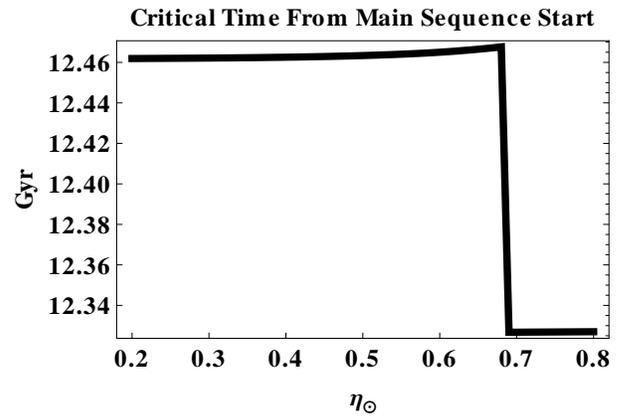,width=8.5cm,height=5.5cm} 
}
\caption{The time from the start of the Sun's main sequence that orbiting bodies
are most likely to escape the Solar System.  Despite the complex variance
in Solar evolution as a function of $\eta_{\odot}$, this critical
time has a relatively smooth, piecewise dependence on $M_{\odot}$.
}
\label{plot5}
\end{figure}

\begin{figure}
\centerline{
\psfig{figure=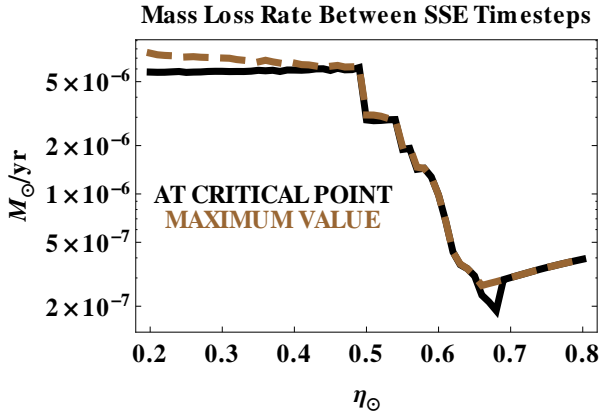,width=8.5cm,height=5.5cm} 
}
\caption{The mass loss rate at the moment when orbiting 
bodies are most likely to escape (solid black
curve) and when this value takes on a maximum
(dashed brown curve).  The excellent agreement between
the two curves suggests that mass loss rate alone,
regardless of the amount of mass remaining in the Sun,
is the primary indicator of when an orbiting body is
no longer safe from ejection.
}
\label{plot6}
\end{figure}

\begin{figure}
\centerline{
\psfig{figure=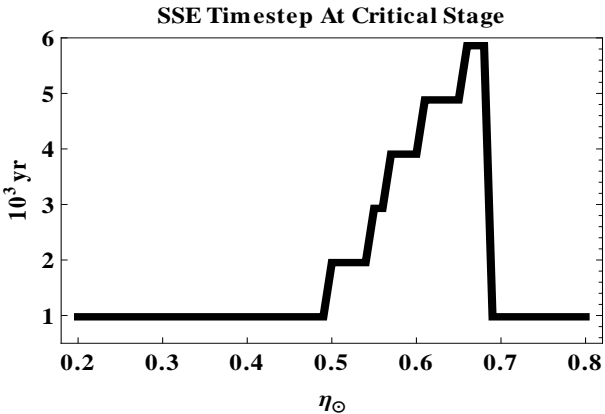,width=8.5cm,height=5.5cm} 
}
\caption{The SSE timestep at the moment when orbiting 
bodies are most likely to escape.  Although short on
the timescale of Solar System evolution, this timestep
is long relative to a Solar rotation period, where additional
variability which is not modeled here might occur.  The mass
loss over this timestep is treated as linear.
}
\label{plot7}
\end{figure}

Here we take a closer look at the critical stage which yields $a_{\rm crit}$ through
Figs. \ref{plot4}-\ref{plot7}.  Figures \ref{plot4}-\ref{plot5} 
demonstrate that the critical stage occurs nearly at the {\it end} of
the Sun's post-main sequence life, when the Sun
will have already lost the majority of its disposable
mass.  The solid black curve in
Fig. \ref{plot4} is a poor proxy of the $a_{\rm crit}$ curve 
from Fig. \ref{plot3}.  Both Figs. \ref{plot4} and \ref{plot5} suggest
that $\eta_{\odot} = 0.7$ represents a bifurcation point
in Solar evolution, which is approximately where the thermally
pulsing AGB stage becomes transitory.  The critical time from the start of
the Sun's main sequence will be $\approx 12.465 \pm 0.005$ Gyr for $\eta_{\odot} \le 0.7$,
and $12.326 \pm 0.0002$ Gyr for $\eta_{\odot} \ge 0.7$.  These
are remarkably well-constrained values relative to the curves in
Fig. \ref{plot3}.

Now consider $\alpha_{i}$ at the critical stage.  Figure \ref{plot6}
shows that this mass loss rate varies between $\approx 2 \times 10^{-7} M_{\odot}$/yr 
and $\approx 5 \times 10^{-6} M_{\odot}$/yr.  Also, this rate closely 
mirrors the maximum rate achieved throughout the simulation,
given by the dashed, brown curve.  Therefore, by itself, ${\rm max}(\alpha_i)$
is an excellent indicator of the stability prospects of an orbiting 
body.  Also, this trend suggests that 
short violent outbursts from planet-hosting stars can jeopardize
the survival of orbiting bodies.  The Solar System is subject to the
same danger even though the Sun is not expected
to experience a violent outburst with 
$\alpha > 5 \times 10^{-6} M_{\odot}$/yr from these models.  If we are underestimating
the variability of the Sun after it turns off of the main sequence,
then orbiting bodies can be in greater danger of escape than we currently expect.

If such variability is on a timescale which is shorter than the timesteps
used in these simulations, then $a_{\rm crit}$ is being overestimated.  
Figure \ref{plot7} illustrates the timestep from which the mass loss rate
at the critical stage was computed.  In all cases, this timestep is between
approximately $980$ yr and $5860$ yr.  The Sun's current rotational period
is less than one month; during a single rotation, prominences may erupt and
sunspots may appear, among other phenomena.  The Sun's (unknown) post-main sequence 
variability will likely be more violent, but on a similar timescale.  If so, then short
bursts of intense mass loss could be important but are not modeled by
the comparatively long SSE timesteps.

\subsection{Orbital Properties}

\subsubsection{Overview}

Here we analyze some representative 
orbits from the N-body simulations.  Because
the timescale for mass loss from the bottom panel
of Fig. \ref{plot1}
is orders of magnitude shorter than the
top panel,
the difference in orbital 
properties is pronounced.  Specifically, for $a_0 > 10^3$ AU
and $\eta_{\odot} < 0.5$, all eccentricity variations 
will often occur during a single orbit.
Alternatively, for $\eta_{\odot} \ge 0.5$, these
variations occur over many orbits.  We consider
one case from the former category and two cases from
the latter in detail.  Additionally, we consider the
eccentricity evolution for one
case in the runaway regime, with $a_0 = 10^5$ AU and
$\eta_{\odot} = 0.2$, and demonstrate here that objects
near their apocenters do remain bound in this case,
as predicted by the linear theory.

\subsubsection{Example Bounded Orbits for $\eta_{\odot} < 0.5$}

\begin{figure*}
\centerline{
\psfig{figure=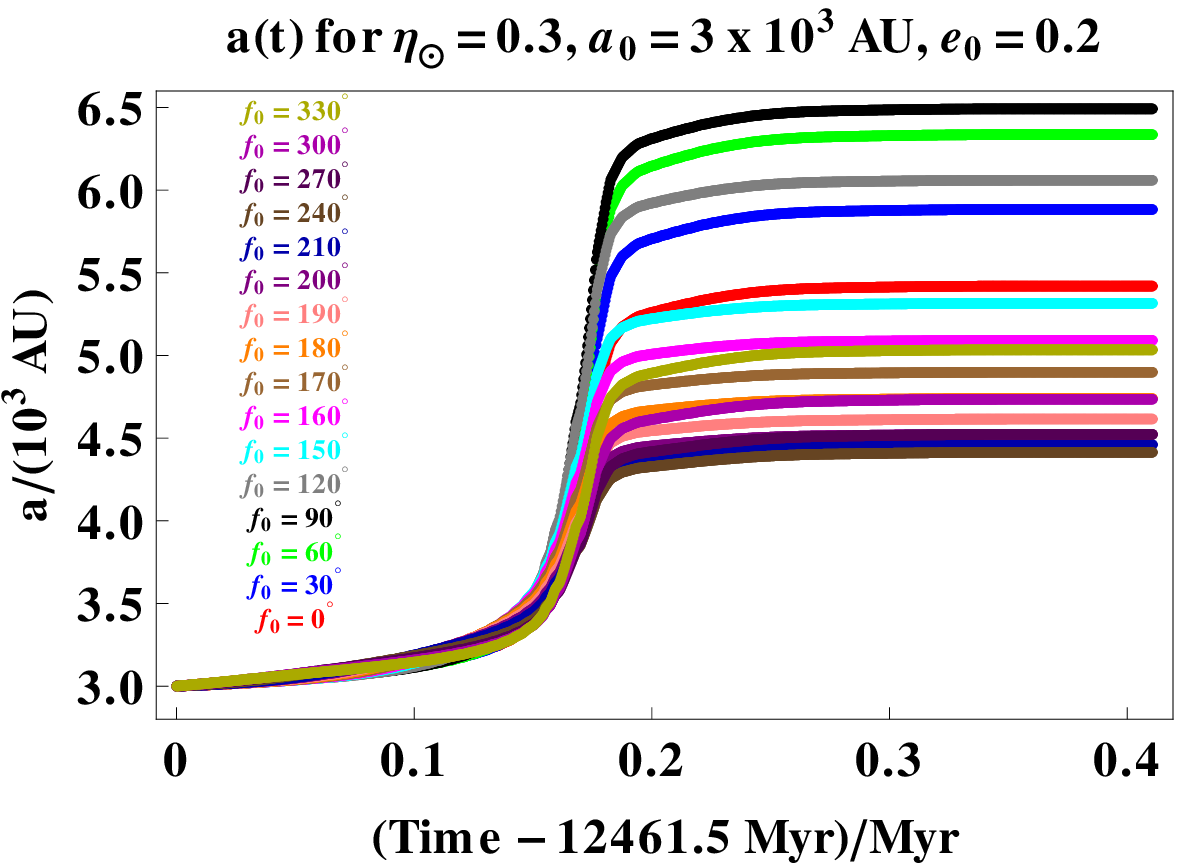,width=8.5cm} 
\psfig{figure=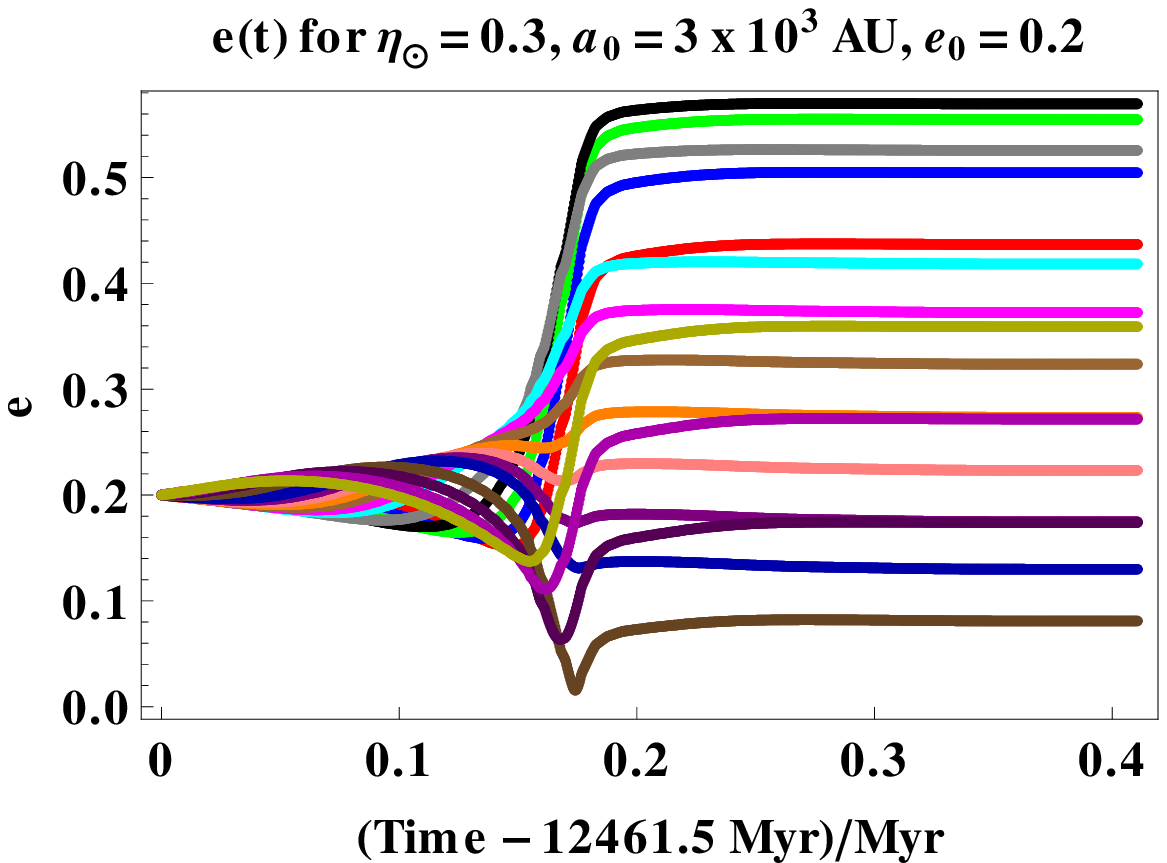,width=8.5cm}
}
\centerline{ }
\centerline{
\psfig{figure=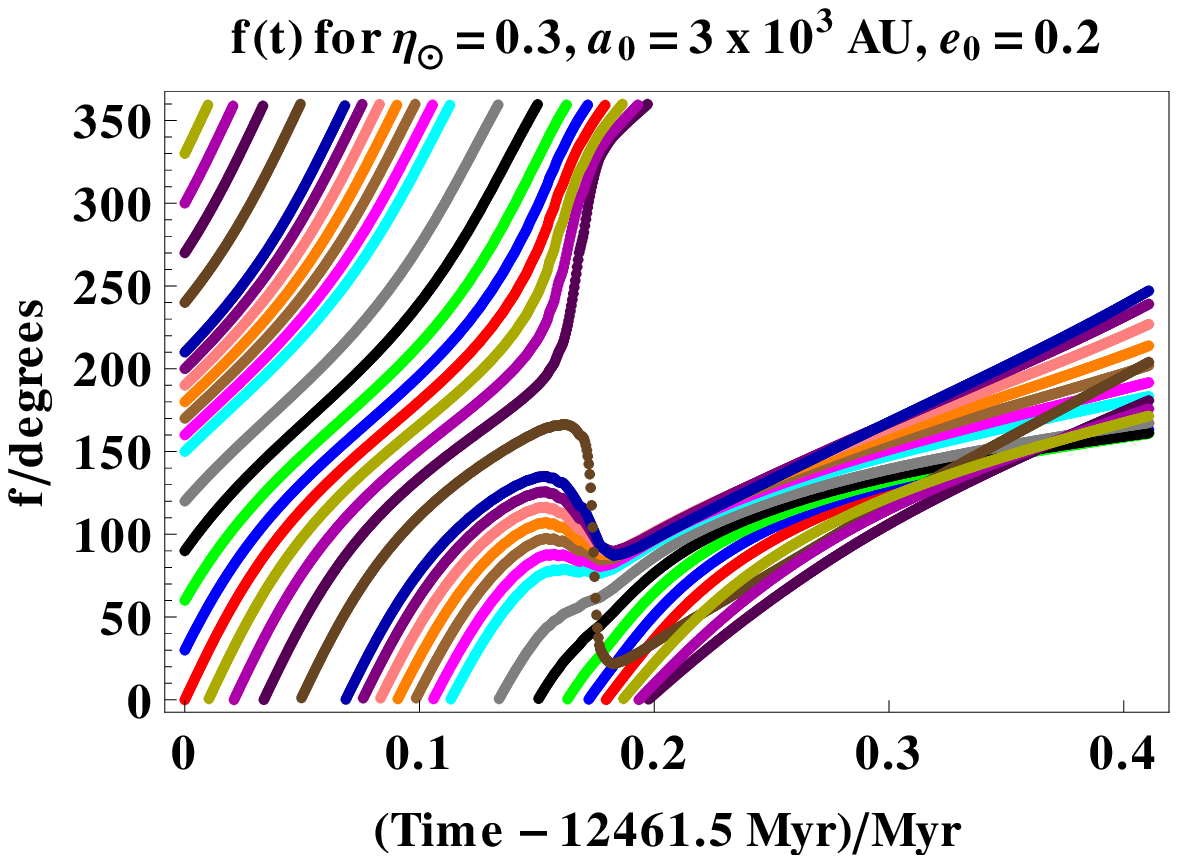,width=8.5cm} 
\psfig{figure=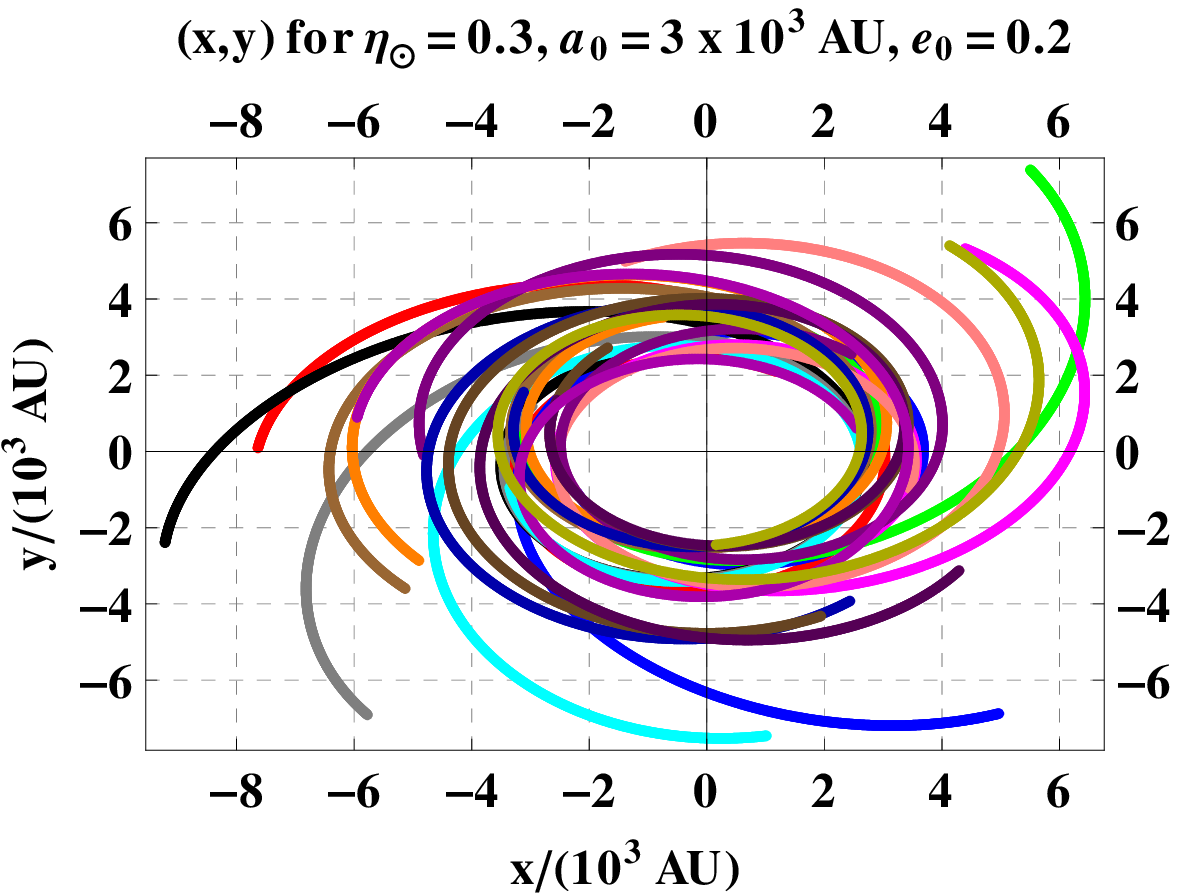,width=8.5cm}
}
\caption{The orbital evolution of objects
at $a_0 = 3 \times 10^3$ AU with $e_0 = 0.2$
from 
$12461.5$ Myr.
All of these bodies
remain bound to the dying Sun, despite sometimes
severe disruption to their orbits; the double cross
in Fig. \ref{plot3} corresponding to this
initial pair $(a_0,\eta_{\odot})$ indicates that escape occurred
for $e_0 \ge 0.8$. The lower-right plot demonstrates 
that the disruption occurs within a single orbital
period, and can stretch or contract orbits (upper-right plot) 
while widening them (upper-left plot).  After this moment in 
time, the orbits assume osculating unchanging ellipses,
and remain on these orbits as the Sun becomes a white dwarf.
The lower-left plot indicates that all objects
barely remain in the adiabatic regime because $df(t)/dt \ne 0$ 
for all cases at all times, 
ensuring that $f$ continues to circulate, albeit slowly.  
The orbit which is most severely disrupted 
($f_0 = 90^{\circ}$, black curve) comes closest to achieving
$df(t)/dt = 0$.
}
\label{plot8}
\end{figure*}

We consider here short orbits ($a_0 = 3 \times 10^3$ AU) 
which are significantly disrupted by mass loss that
takes place entirely within an orbital period 
($\eta_{\odot} = 0.3$, see Fig. \ref{plot1})
and remain ellipses.  Figure \ref{plot8} illustrates the
orbital evolution of 16 objects, all with different $f_0$ values,
which all begin the N-body simulation with small-eccentricity 
($e_0 = 0.2$) orbits.  These objects may be treated as 
low-eccentricity, inner Oort Cloud bodies.
Despite the small $e_0$, the semimajor axes all increase
by a factor of $\approx 1.5-2.2$, and the final values
of $e$ range from $\approx 0.1-0.6$.  If this mass was
lost adiabatically, then the semimajor axis would increase
by a factor of $\approx 1.5$, the lower bound of the actual
increase in $a$.  Further, the
$f_0 = 240^{\circ}$ object (brown curves) becomes nearly circularized at 
$t \approx 12461.67$ Myr; if mass loss ceased at this point,
then the body would maintain this nearly circular orbit.

Mass loss has a dramatic effect on the evolution of $f$.
A system no longer is adiabatic when $df/dt = 0$
\citep{veretal2011}.  At this
point, $f$ stops circulating and begins to librate.
The lower left plot illustrates how the mass loss
slows $df/dt$ for each orbiting object.  Although none
of the orbiting objects achieve $df/dt = 0$,
the $f_0 = 90^{\circ}$ object (black curves)
comes closest.  This is also the object
whose $a$ and $e$ values are increased by the greatest
amount.  Alternatively, the highest ending 
value of $df/dt$ is associated with the $f_0 = 240^{\circ}$ object, 
whose eccentricity decreases the most. 
In the linear mass loss approximation, for $\Psi \gg 1$, bodies closest
to pericenter are most susceptible to escape.  This plot
helps demonstrate how no such correlation exists if the
evolving regime is not runaway.

Although the bottom-right plot might
suggest that some planets appear to be leaving this system,
all bodies will remain bound, orbiting the Solar
white dwarf with their values of $a$ and $e$ attained
at $t = 12461.9$ Myr, assuming no additional perturbations.

\subsubsection{Example Escape Orbits for $\eta_{\odot} \ge 0.5$}

\begin{figure*}
\centerline{
\psfig{figure=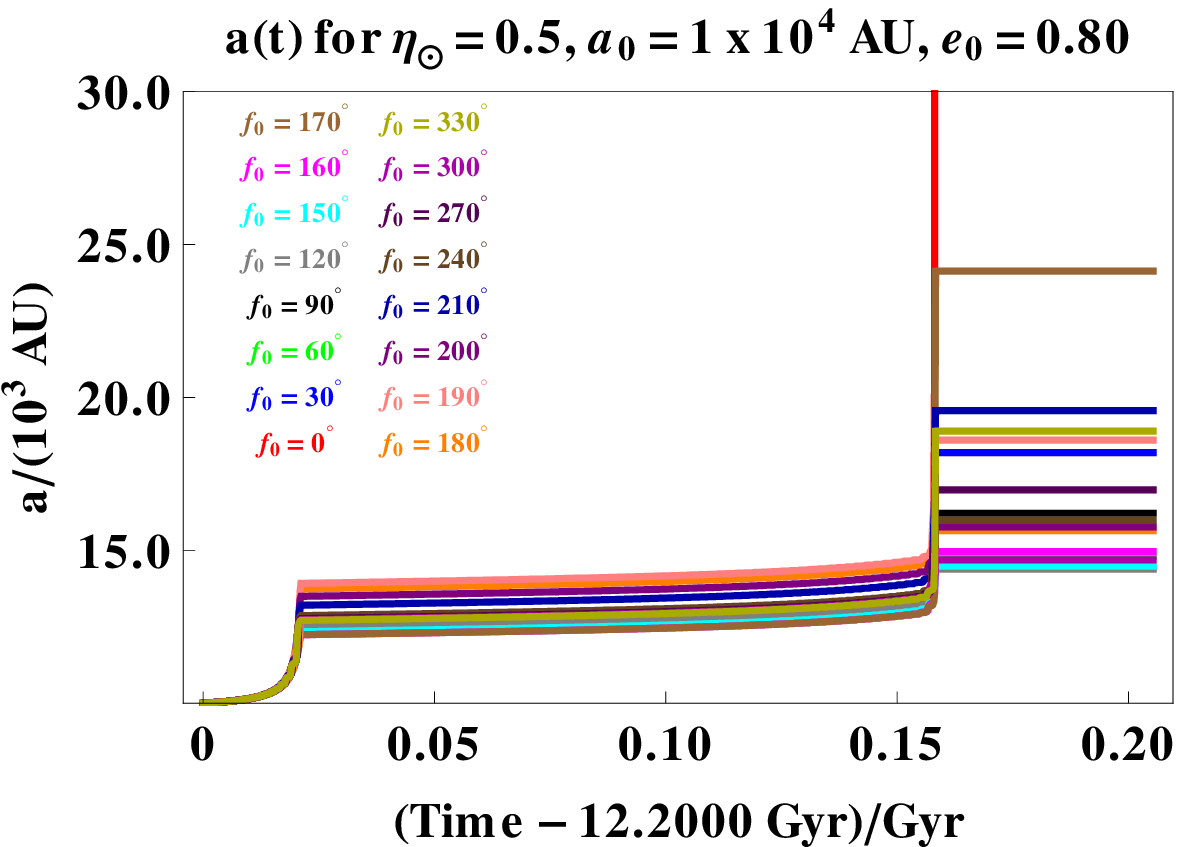,width=8.5cm,height=4.5cm} 
\psfig{figure=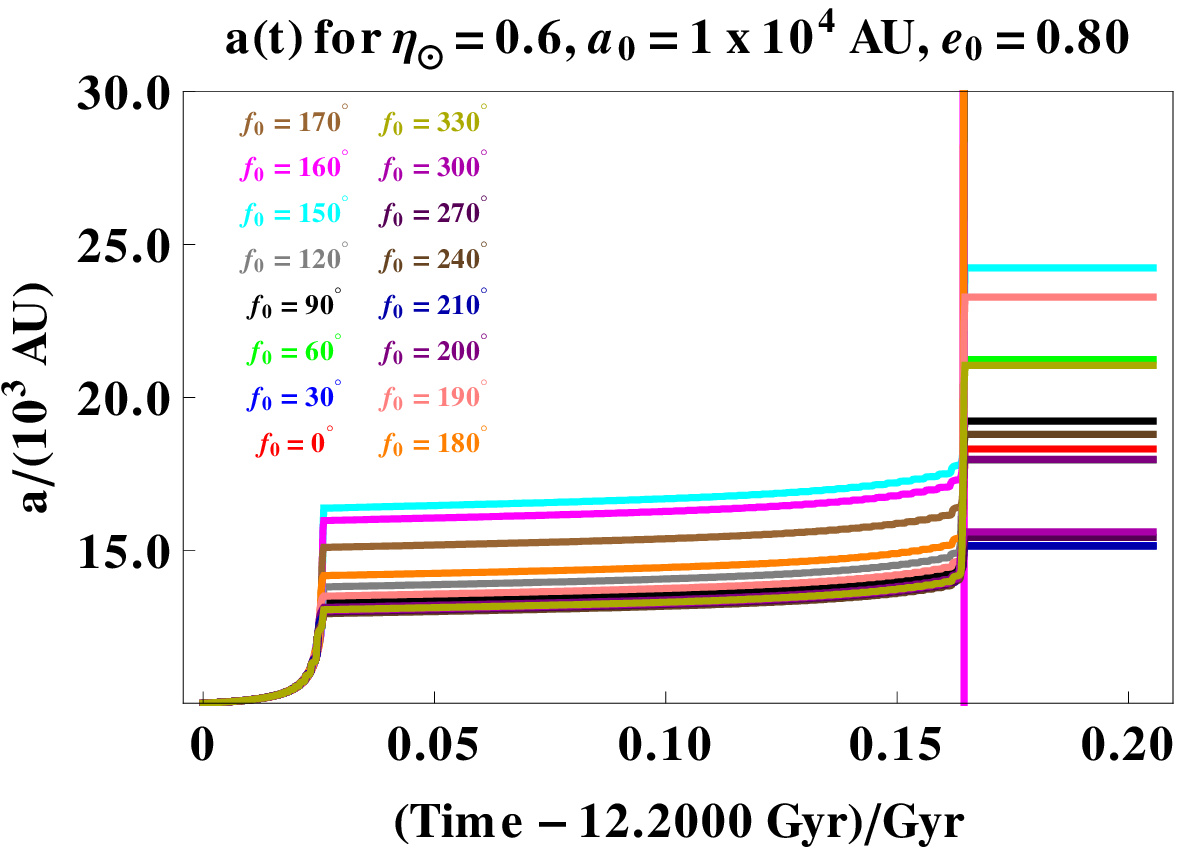,width=8.5cm,height=4.5cm}
}
\centerline{ }
\centerline{
\psfig{figure=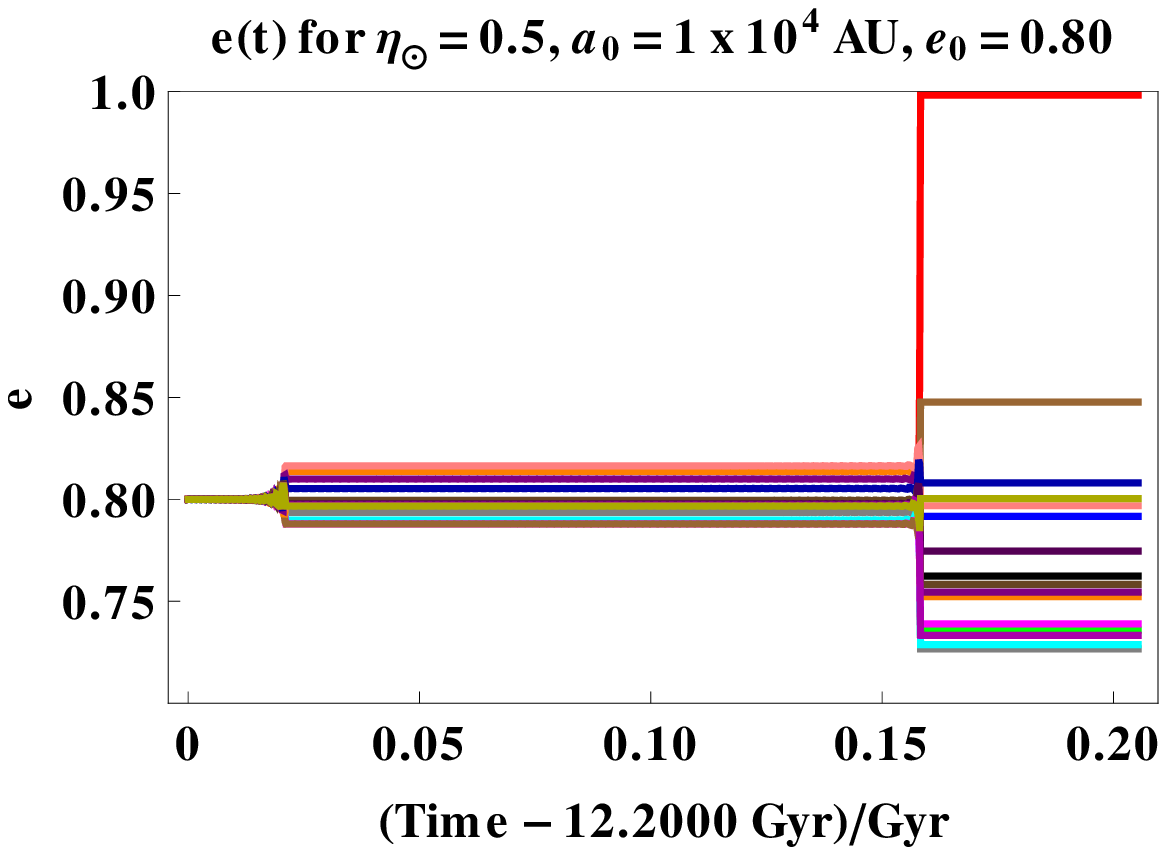,width=8.5cm,height=4.5cm} 
\psfig{figure=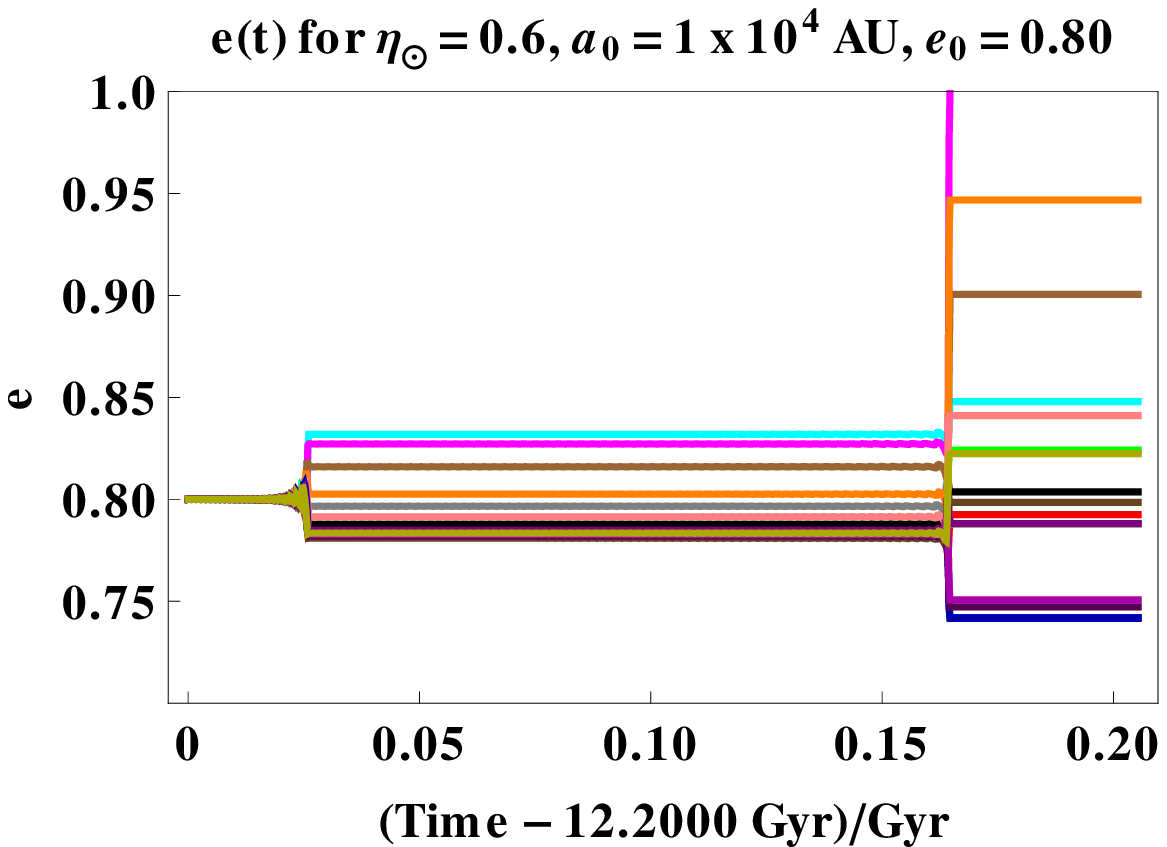,width=8.5cm,height=4.5cm}
}
\centerline{ }
\centerline{
\psfig{figure=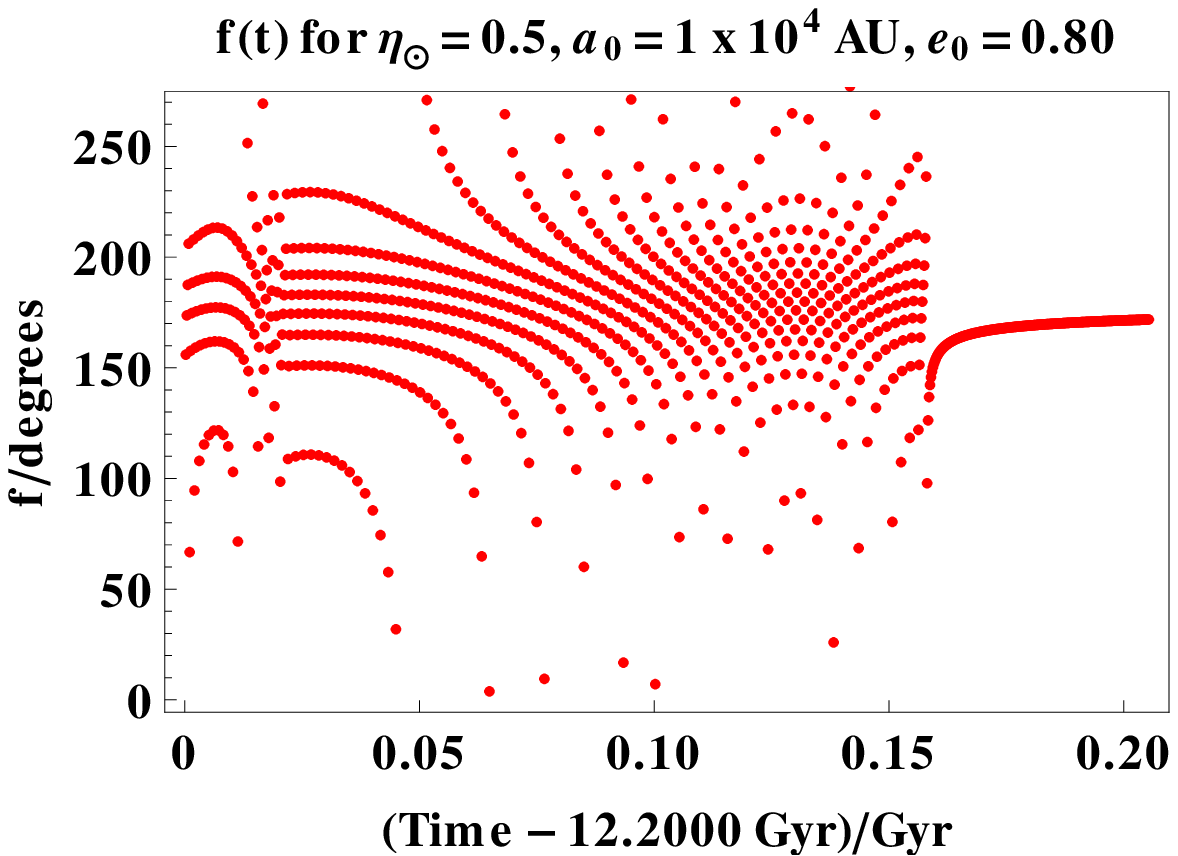,width=8.5cm,height=4.5cm} 
\psfig{figure=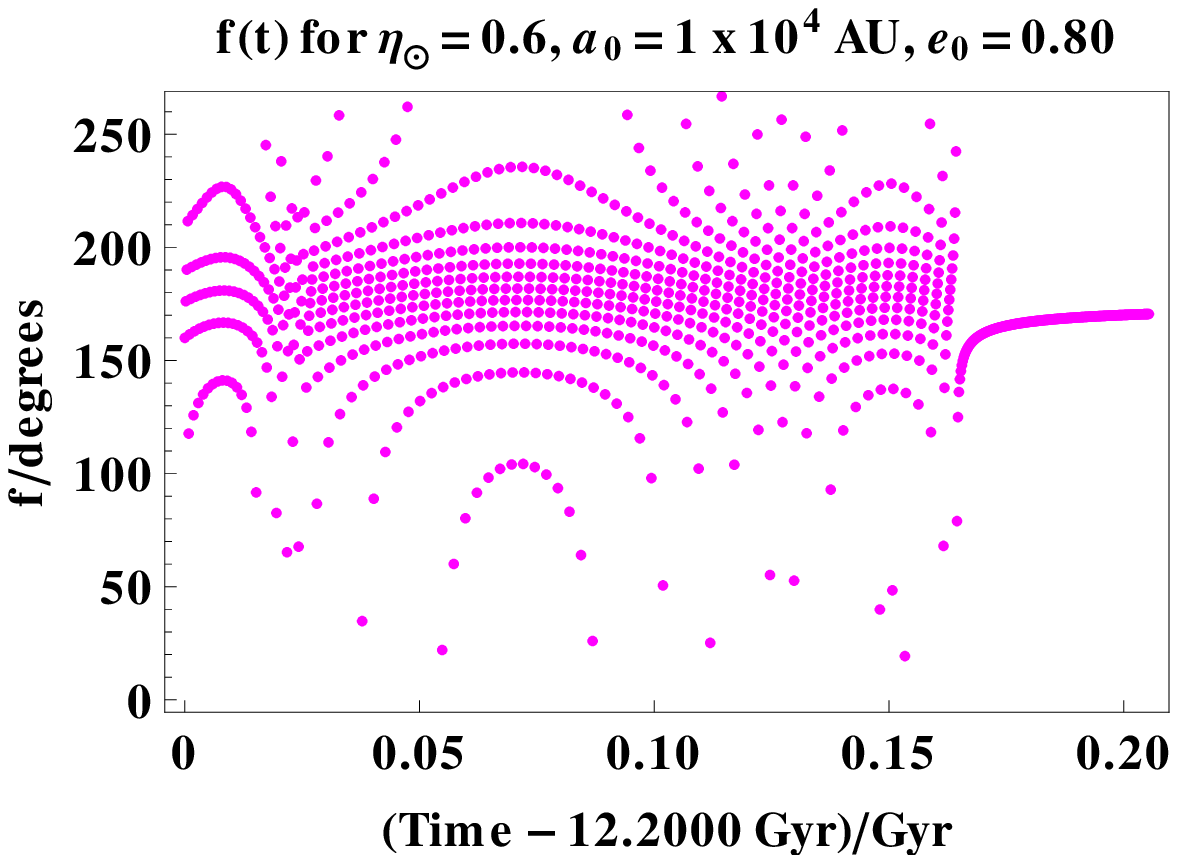,width=8.5cm,height=4.5cm}
}
\centerline{ }
\centerline{
\psfig{figure=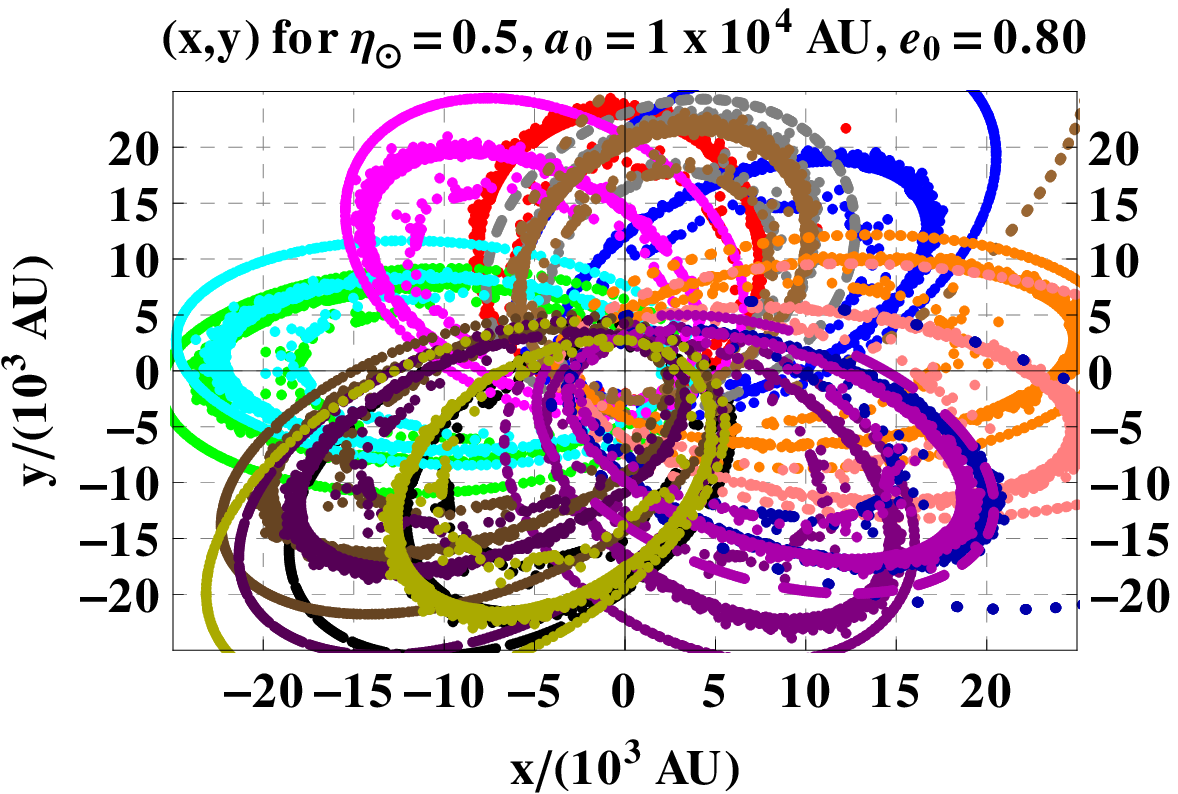,width=8.5cm,height=4.5cm} 
\psfig{figure=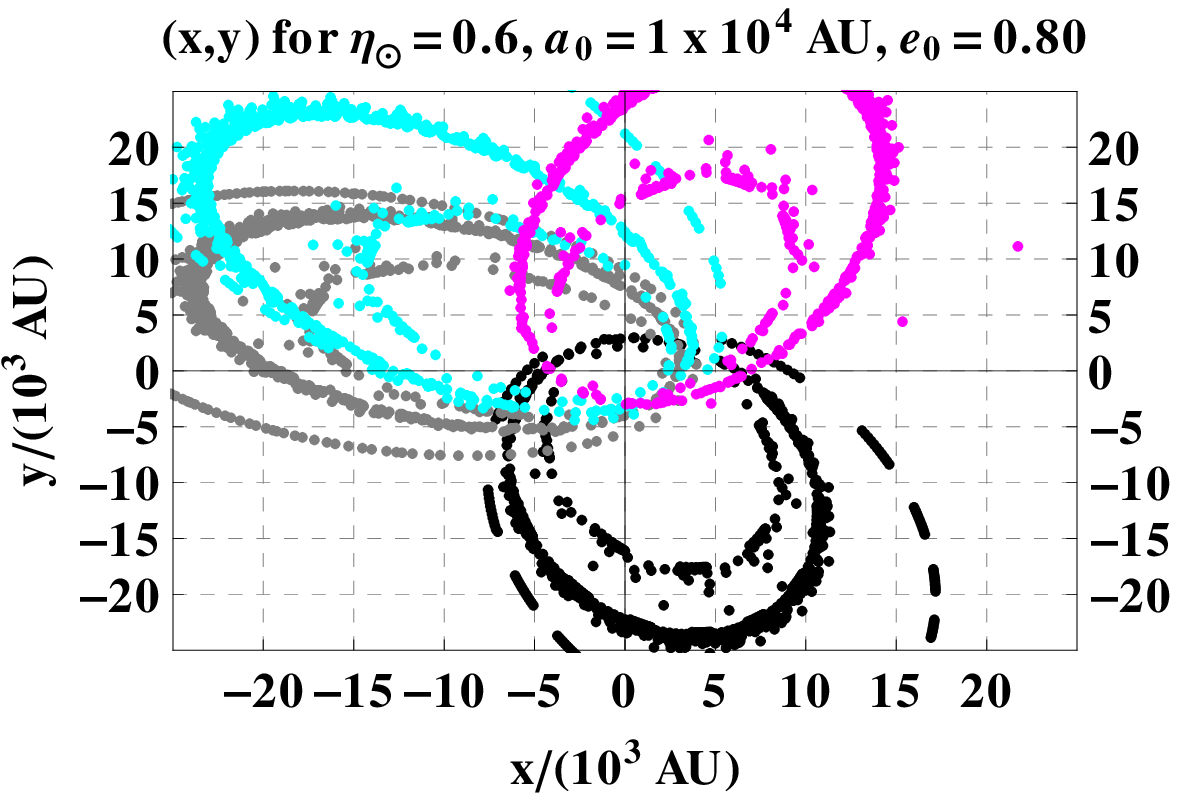,width=8.5cm,height=4.5cm}
}
\caption{The orbital evolution of objects
at $a_0 = 1 \times 10^4$ AU with $e_0 = 0.8$
for $\eta_{\odot} = 0.5$ (left panels) and
$\eta_{\odot} = 0.6$ (right panels) from 
$12.2$ Gyr.
In the left
panels, the $f_0 = 0^{\circ}$ (red) body escapes
when, after achieving $e = 0.99872$, its semimajor
axis exceeds $10^6$ AU.  Alternatively, in the right panels, the
$f_0 = 160^{\circ}$ (pink) body escapes by achieving
a hyperbolic orbit ($e > 1$, $a < 0$).  The $f$ vs $t$ panels
demonstrate that the true anomaly cannot distinguish
between these two types of escape in this case, and
that bodies may escape just before achieving 
$df(t)/dt = 0$.  The pericentric vs. near-apocentric
$f_0$ values of the 
escaping bodies, as well as well as a comparison
of the orbit locations in the bottom two plots,
indicate that the phase information is easily lost
on the approach to $df(t)/dt = 0$. The bottom plots also
demonstrate how the orbital architecture changes
over several orbital periods, in contrast to Fig. \ref{plot8}.
Note that the ejected planet in the bottom-right
plot never comes within its initial pericenter value ($2000$ AU)
of the Sun.
}
\label{plot9}
\end{figure*}

Here we consider longer orbits ($a_0 = 1 \times 10^4$ AU) 
which are significantly disrupted by mass loss that
takes place over several orbital periods 
($\eta_{\odot} = 0.5, 0.6$, see Fig. \ref{plot1})
and which can cause an orbiting object to escape.  We
consider highly eccentric orbits ($e_0 = 0.8$)
and compare two types of escape in Fig. \ref{plot9}:
the left panels showcase the $f_0 = 0^{\circ}$ (red)
object achieving an orbit with $e = 0.99872$
and $a > 10^6$ AU (two orders of magnitude 
higher than $a_0$), while the right panels
showcase the $f_0 = 160^{\circ}$ (pink)
object achieving a hyperbolic orbit.
The only initial parameter changed in the two
plot columns was the value of $\eta_{\odot}$ 
(0.5 on the left, and 0.6 on the right).

As evidenced by the plots in the upper two
rows, the first strong burst of mass loss
at $t \approx 12.22$ Gyr 
provides a stronger orbital kick for $\eta_{\odot} = 0.6$
Solar evolution.  However, as indicated
by Fig. \ref{plot3}, the second, stronger
burst of mass loss at $t \approx 12.36$ Gyr
yields a lower value of $a_{\rm crit}$ for
$\eta_{\odot} = 0.5$.  Regardless, this second
strong burst of mass loss for both $\eta_{\odot}$ values
are roughly comparable, as the final range
of $a$ and $e$ values are similar.

The difference
is enough, however, to cause escape in two different
ways, one with an object which was at pericenter
at the beginning of the simulation, and one with
an object which was near-apocenter.  This figure
provides further confirmation of the difficulty with
correlating $f_0$ with $f$ for objects
evolving in the transitional regime between
the adiabatic and runaway regimes.  Both
$f$ vs $t$ plots demonstrate the expected change
of behavior during the second strong mass loss event.
However, they fail to distinguish between both
types of escape, and neither technically reach $df/dt = 0$.
However, for the left and right panels, respectively,
$df/dt = 0.068^{\circ}/$Myr and $df/dt = 0.078^{\circ}/$Myr
at $t = 12461.9$ Myr.

The bottom panels show the spatial orbits of the objects;
the left panel illustrates all 16, and the right panel
features only 4 (for added clarity), including the escaping object.
Due to the number of simulation outputs, lines are not connected
between the outputted points.
These plots explicitly illustrate how two bursts
of mass loss from the Sun cause two distinct orbit expansions.
Additionally, the apparent ``noise'' about the equilibrium orbits
attained before the second mass loss event demonstrate
the gradual expansion of the orbit due to minor mass loss events.
Comparing the colours and positions of both plots shows
almost no correlation with the identical
initial conditions.  The escaping pink body on the right
panel never reaches a distance to the Sun which 
is closer than its initial pericenter value ($2000$ AU);
the two dots for which $x > 1.5 \times 10^4$ AU
indicate the path along which the body escapes.

\subsubsection{Example Orbits for Rampant Escape}

We now consider objects in considerable
danger of escape.  These {\it can} reside at typical
Oort Cloud distances, $a_0 \sim 10^5$ AU.
The strong mass loss from the $\eta_{\odot} = 0.2$
evolutionary track (with maximum 
$\alpha \approx 7.5 \times 10^{-6} M_{\odot}$/yr; see Fig. \ref{plot6})  
yields $\Psi \gtrsim 30 \gg 1$ and places these objects
in the runaway regime.  Here we determine if, unlike
in the adiabatic regime, we can link the fate of an object
with $f_0$ amidst nonlinear mass loss.

Figure \ref{plot10} confirms that we can.  The results
agree well with the linear theory.  
At the highest eccentricities sampled ($e_0 = 0.95$), the
objects initially at apocenter (orange; $f_0 = 180^{\circ}$) 
remain bound to the dying Sun, and, further, suffer the
greatest eccentricity decrease.
Additionally, 4 other objects nearest to the apocenter of their
orbits also remain bound.  As the initial eccentricity of
the objects is decreased (from $0.95$ in the top panel
to $0.6$ in the bottom panel), more bodies which are increasingly
further away from their initial apocenter are allowed to survive.
Also, as the initial eccentricity is decreased, more bodies which
are progressively further away from their initial apocenter
suffer a net eccentricity decrease.  Note finally that the
results are not symmetric about $f_0 = 180^{\circ}$: objects
approaching apocenter are slightly more protected than
objects leaving apocenter.

\section{Application to the Scattered Disc, Oort Cloud and other Trans-Neptunian Objects}

\subsection{Overview}

The Solar System architecture beyond the Kuiper Belt 
is often divided into two distinct regions: the scattered 
disc and the Oort Cloud.  The former is possibly an ancient, eroding remnant
of Solar System formation, whereas the latter is thought to dynamically interact
with the local Galactic environment and is continuously 
depleted and replenished.  Both regions may supply each other with
mass throughout the lifetime of the Solar System.

However, the definitions of these regions are not standardized.  
The boundary of the scattered disc is empirical
and refers to objects with perihelia beyond Neptune ($q > 30$ AU)
and semimajor axes beyond the $2$:$1$ resonance with Neptune ($a > 50$ AU)
\citep{luetal1997,gometal2008}.  Other papers have set a variety
of numerical bounds based on their empirical readings of
the semimajor axis - eccentricity phase space for objects beyond Neptune.
In an early study, \cite{dunlev1997} distinguished 
the Kuiper Belt from a scattered disc
of objects with $q \approx 32-48$ AU, $e \le 0.8$ and $i \le 50^{\circ}$.
Alternatively, sometimes the scattered disc is considered
to be a subclass of the Kuiper Belt \citep[e.g.][]{glacha2006} and contains objects which 
satisfy the simpler constraint $q \gtrsim 33$ AU \citep{tismal2003,volmal2008} and 
$e \gtrsim 0.34$ \citep{volmal2008}, whereas \cite{kaiqui2008} give $q \ge 40$ AU 
and \cite{sheppard2010} claim $q \sim 25-35$ AU
as the criterion for inclusion.  
Further, \cite{levetal2004} demonstrate how the scattered disc population
has changed with time.  For example, at an age of $10^6$ yrs, most scattered disc objects 
had $q < 10$ AU.  More generally, \cite{levetal2008} define 
the currently-observed scattered disc to be composed of objects with perihelia 
close enough to Neptune's orbit to become unstable over the (presumably 
main-sequence) age of the Solar System.  However, some scattered disc objects
may not be primordial, and instead represent a small transient population supplied
by the Kuiper Belt.  Also, because scattered disc objects interact with Neptune, some
are depleted and diffuse into the Oort cloud.

If the scattered disc is insufficiently replenished from the Oort Cloud and Kuiper Belt, 
then the scattered disc 
may be depleted relative to its current level by the end of the Sun's main sequence.
Whatever disc objects which might remain will continue
to interact with Neptune as their orbits move outward.  If
the orbits expand adiabatically, the scattered disc object will continue 
to interact with Neptune as during the main sequence.  
If the orbits expand non-adiabatically, then the scattered disc objects'
eccentricity will change, and may no longer interact with Neptune.   
However, the semimajor axis of a scattered disc object which is high enough to cause
non-adiabatic evolution is generally too high ($a > 10^3$ AU) to remain in that population.
Therefore, the scattered disc population will be subject to adiabatic mass
loss evolution, and continue to interact by diffusing chaotically in semimajor axis \citep{yabushita1980}, likely
helping to repopulate the Oort Cloud.

\begin{figure}
\centerline{
\psfig{figure=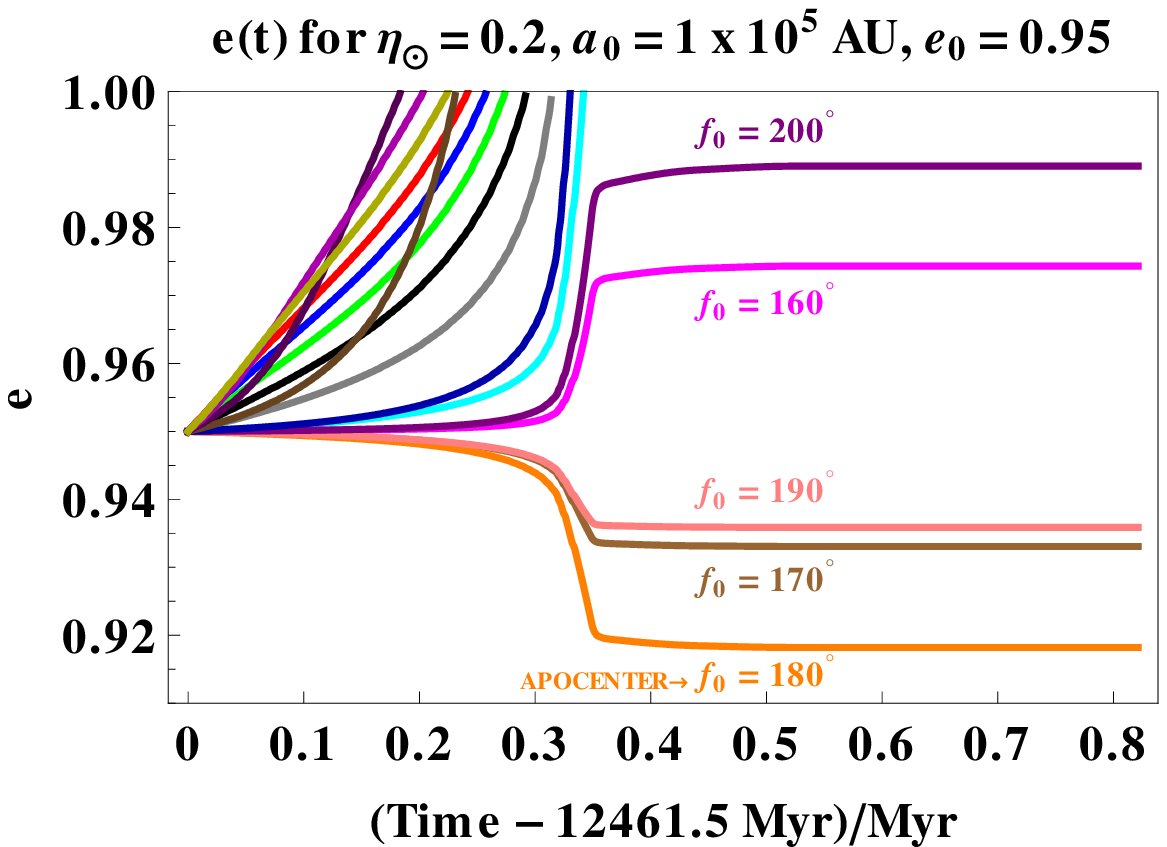,width=8.5cm}
}
\centerline{ }
\centerline{
\psfig{figure=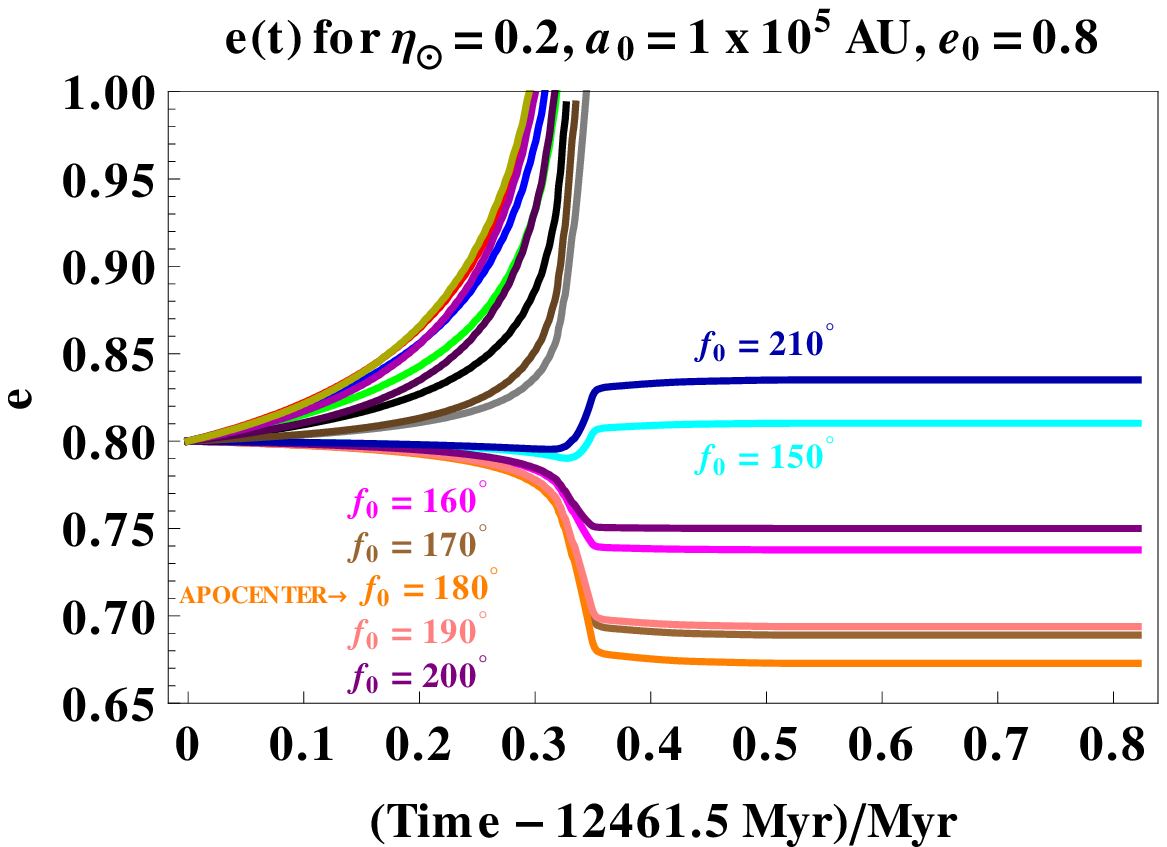,width=8.5cm}
}
\centerline{ }
\centerline{
\psfig{figure=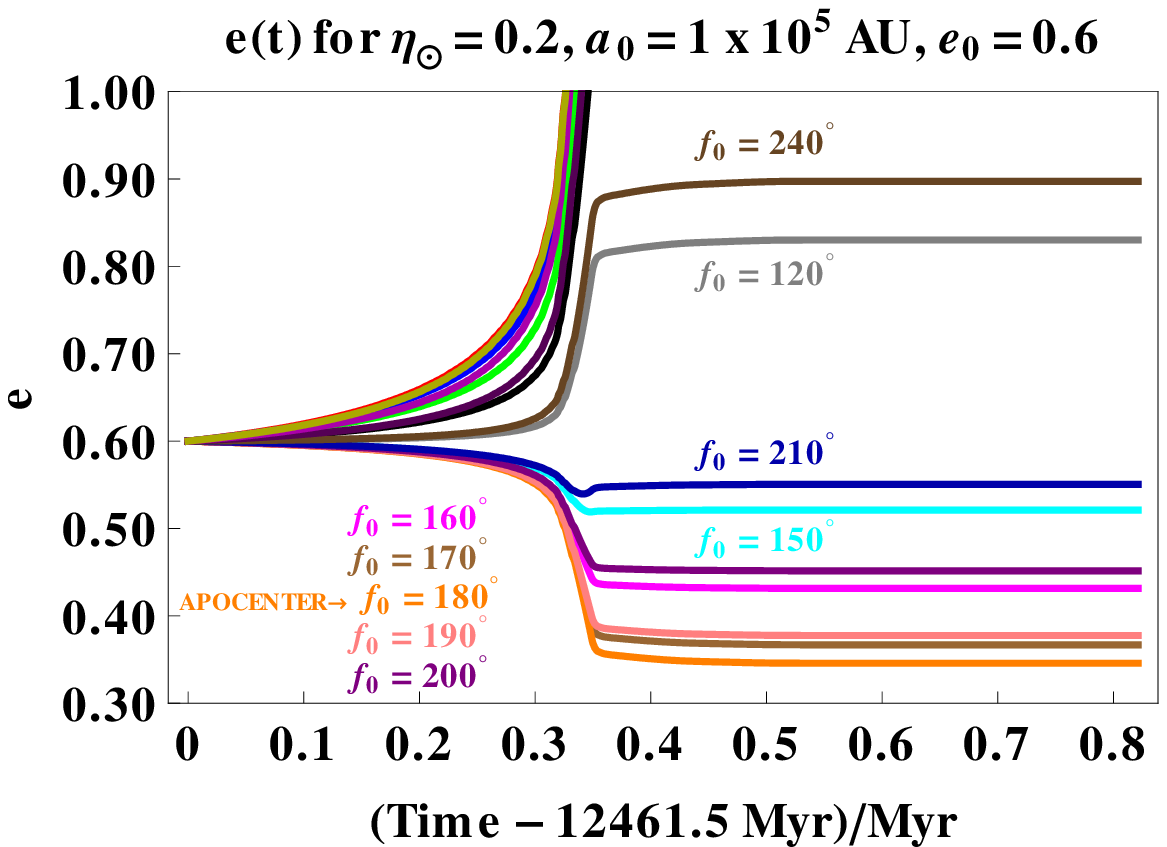,width=8.5cm}
}
\caption{Eccentricity evolution in the runaway regime for typical
Oort Cloud distances of $10^5$ AU and $\eta_{\odot} = 0.2$. 
Objects initially closest
to the apocenter of their orbit are the most likely to survive,
and bodies at their apocenters experience the largest net drop
in eccentricity.  Objects further away from the apocenter
may survive if their initial eccentricity is decreased (from
$0.95$ in the top panel to $0.6$ in the bottom panel).
}
\label{plot10}
\end{figure}

Beyond the scattered disc resides the Oort Cloud, which was originally postulated to 
contain $\sim 10^{11}$ comets of observable size 
with $5 \times 10^4$ AU $< a < 1.5 \times 10^5$ AU, $0 < e < 1$ and isotropically-distributed 
inclinations \citep{oort1950}.  Some of these comets 
occasionally enter the inner Solar System and can achieve perihelia of a few AU.
These intruders are thought to have been jostled inward by forces external to the Solar System.
Many subsequent investigations since this seminal work have refined this basic picture for 
the outermost region of the Solar System.  

The external perturbations may arise from passing stars and/or the local Galactic tide.  \cite{heitre1986}
demonstrated that tides likely dominate injection events into the inner Solar System,
while passing stars continuously randomize the Oort Cloud comet 
orbital parameter distribution \citep{dyb2002}.  \cite{mormul1986} demonstrate how Galactic
forces vary the orbital angular momentum of an Oort Cloud comet during a single orbit, 
which, hence, may be neither an ellipse nor closed.  Therefore, if a typical comet's bound orbit 
significantly varies from Keplerian motion due to external perturbations, the 
analytics in \cite{veretal2011} cannot be applied to these objects without incorporating 
models of those perturbations.

However, if we do treat bound Oort Cloud cometary orbits as approximately elliptical, then we can 
estimate the effect of stellar mass loss on the cloud.  One major refinement of Oort's
original model is the division of the Oort Cloud into an ``inner'' and ``outer'' population 
at $a = 2 \times 10^4$ AU \citep{hills1981}.  This bifurcation point, which determines
if a comet is observable only when it is part of a shower, is still widely utilized
in modern simulations and computations.  Figure \ref{plot3} indicates that this bifurcation value 
is a factor of $2-20$ times as high as the Solar System's post main-sequence escape boundary.  
Hence, comets from both populations
will be susceptible to escape due to post main-sequence mass loss.  Moreover, in a seminal
study, \cite{dunetal1987} approximated the inner edge of the Oort Cloud to 
be $\approx 3000$ AU, whereas \cite{brasser2008}
use this same value to divide the Oort Cloud further into a third ``innermost'' region.
High-eccentricity bodies with semimajor axes approximately equal to this potential 
second bifurcation point in Oort Cloud 
structure may be susceptible to escape during the Sun's post-main sequence 
evolution if $\eta_{\odot} < 0.6$ (see Fig. \ref{plot3}).

Many modern Oort Cloud investigations focus on formation and/or the orbital properties of observed
comets instead of topics more relevant to this study, such as the eccentricity distribution 
of the vast majority of the remaining (currently unobservable) comets and the future evolution of the Oort
Cloud over the next several Gyr.  However, some authors \citep{emeetal2007,braetal2010}
present plots which illustrate that the eccentricity distribution of Oort Cloud comets do span from 0 to 1,
with the minimum eccentricity typically increasing as $a$ decreases.  Therefore, the orbits studied
in Sections 3.6.2 and 3.6.3 represent plausible families of ``inner'' Oort Cloud orbits,
and the orbits presented in Section 3.6.4 represent a plausible family of ``outer'' Oort Cloud orbits.

Multiple subpopulations of the Oort Cloud help categorize recent discoveries.  
Sedna \citep{broetal2004}, the most distant Solar System object 
yet observed (at $90.32$ AU), 
highlights the failure of the traditional scattered disc and Oort Cloud to partition
the Solar System beyond the Kuiper Belt.  With $q \approx 76$ AU and $a \approx 480$ AU, Sedna 
does not fit inside either population, and demonstrates that the orbital parameter phase 
space of trans-Neptunian objects is larger than previously thought.  If Sedna 
maintains its orbit until the Sun turns off of the main sequence, then the object is guaranteed
to remain bound (Fig. \ref{plot3}) during post-main sequence mass loss, regardless of
its position along its orbit.  However, other objects a few hundred AU more distant
might not survive.  Two other objects which defy easy classification are 2006 SQ$_{372}$ \citep{kaietal2009}, 
with $q \approx 24$ AU and $a \approx 796$ AU, and 2000 OO$_{67}$ \citep{veietal2001,miletal2002}, with 
$q \approx 21$ AU and $a \approx 552$ AU.  The perihelions and aphelions of both objects
imply that they can be classified as either scattered
disc objects or Oort Cloud comets.  Regardless, the short stability timescale ($\sim 200$ Myr) predicted for 
2006 SQ$_{372}$ implies that it will not survive the Sun's main sequence phase.  2000 OO$_{67}$ will
remain bound (Fig. \ref{plot3}) to the post-main sequence Sun if the object survives the Sun's main sequence.

\subsection{Depletion Characteristics}

Although determining the fraction of the Oort Cloud which is depleted likely requires detailed modeling and
extensive N-body simulations, we can provide some estimates here.  We perform an additional 3 sets
of simulations, for $\eta_{\odot} = 0.2, 0.5$ and $0.8$, and focus on the region beyond $10^4$ AU, where
most Oort Cloud objects are thought to reside.  These values of $\eta_{\odot}$ represent, in a sense, one
nominal and two extreme Solar evolutionary tracks.  For each $\eta_{\odot}$, we integrate with exactly the same
conditions as described in Section 3.3, but now sample $72$ uniformly-spaced values of $f_0$ (for $0.5^{\circ}$
resolution), 9 uniformly-spaced values of $e_0$ in the range $[0.1-0.9]$, and the following 6
values of $a_0/$AU$ = 1.0 \times 10^4, 2.5 \times 10^4, 5.0 \times 10^4, 7.5 \times 10^4, 1.0 \times 10^5, 1.25 \times 10^5$.

\begin{figure}
\centerline{
\psfig{figure=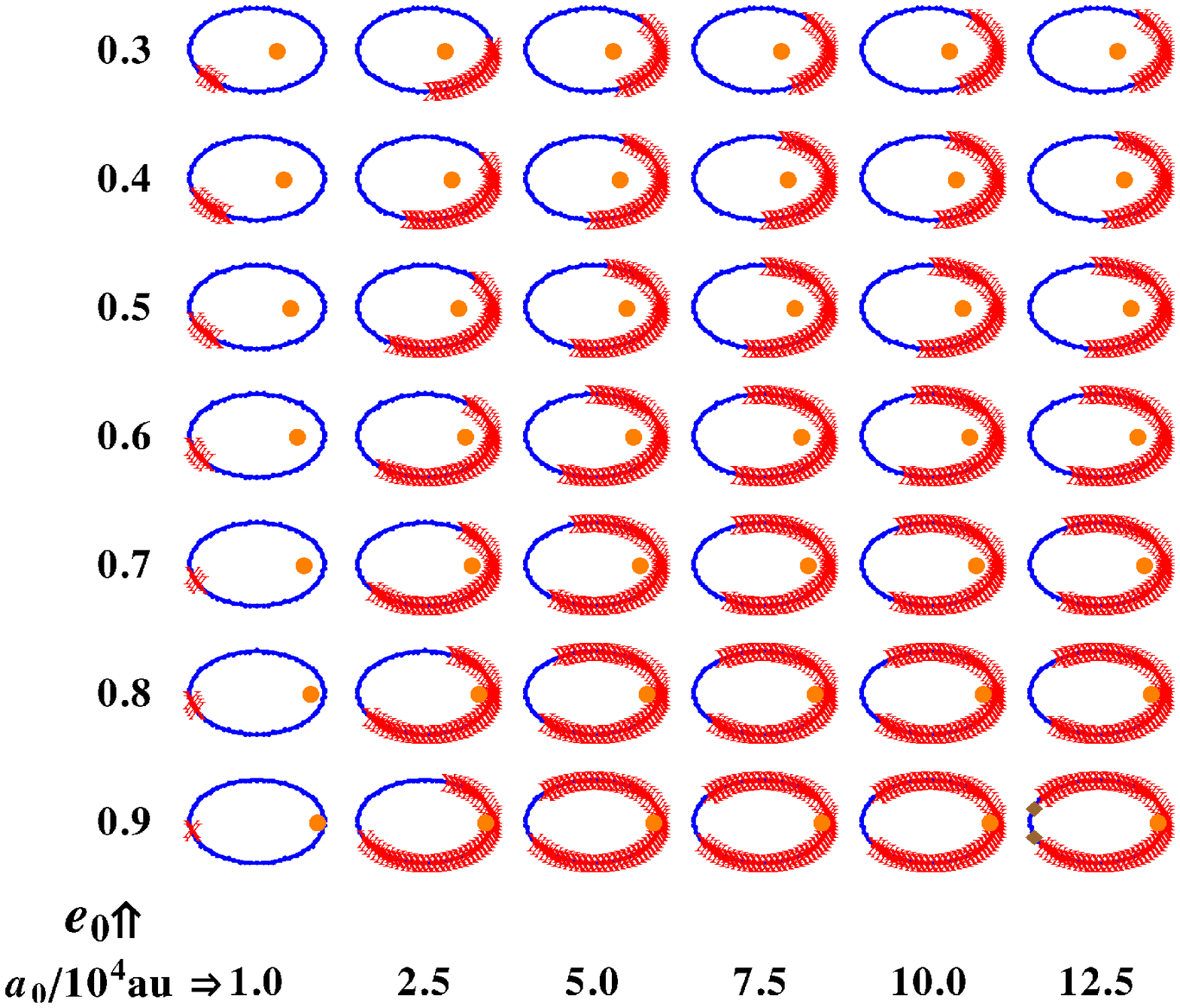,width=8.5cm,height=5.7cm}
}
\centerline{$\boldsymbol{\eta_{\odot} = 0.2}$}
\centerline{
\psfig{figure=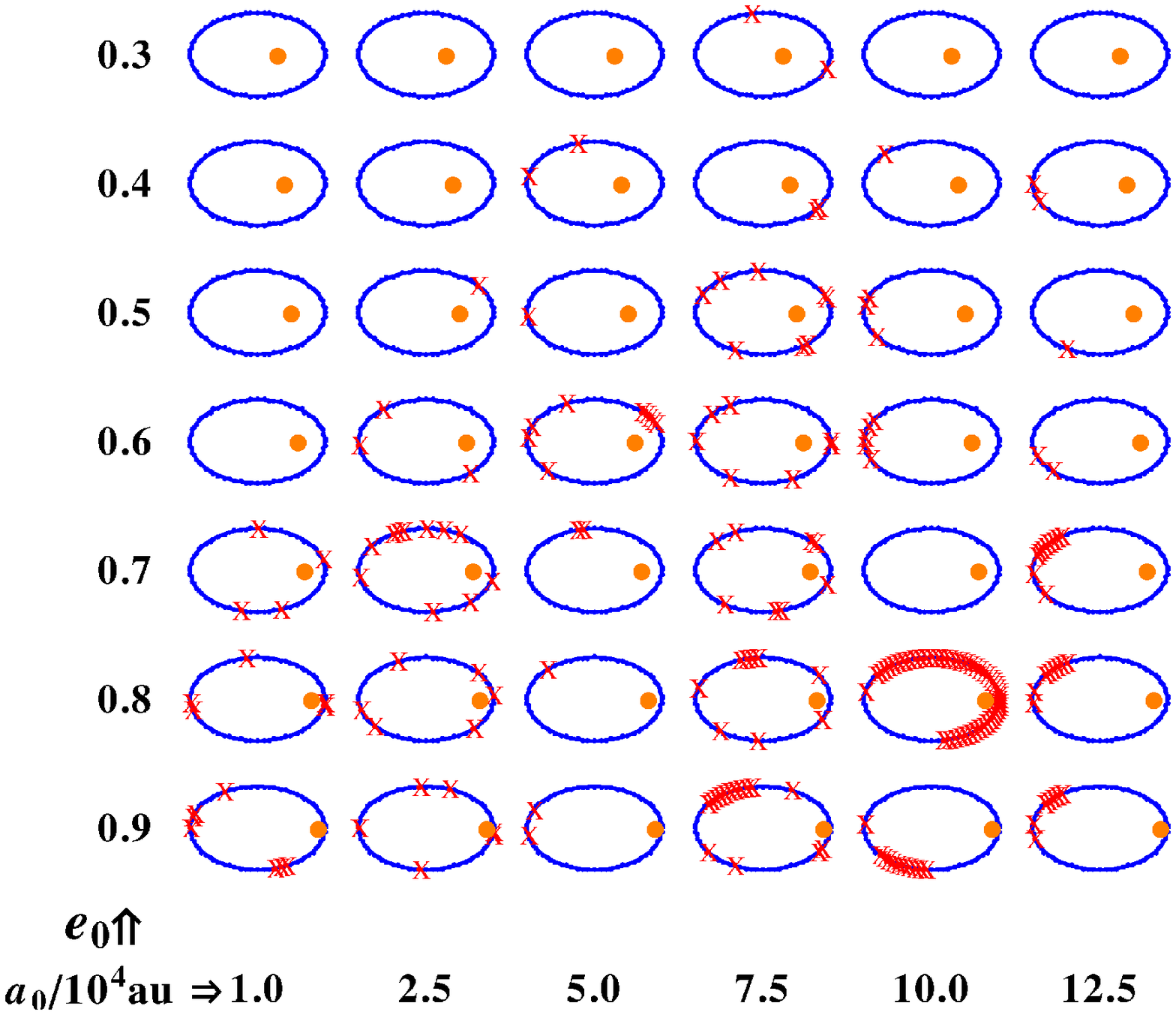,width=8.5cm,height=5.7cm}
}
\centerline{$\boldsymbol{\eta_{\odot} = 0.5}$}
\centerline{
\psfig{figure=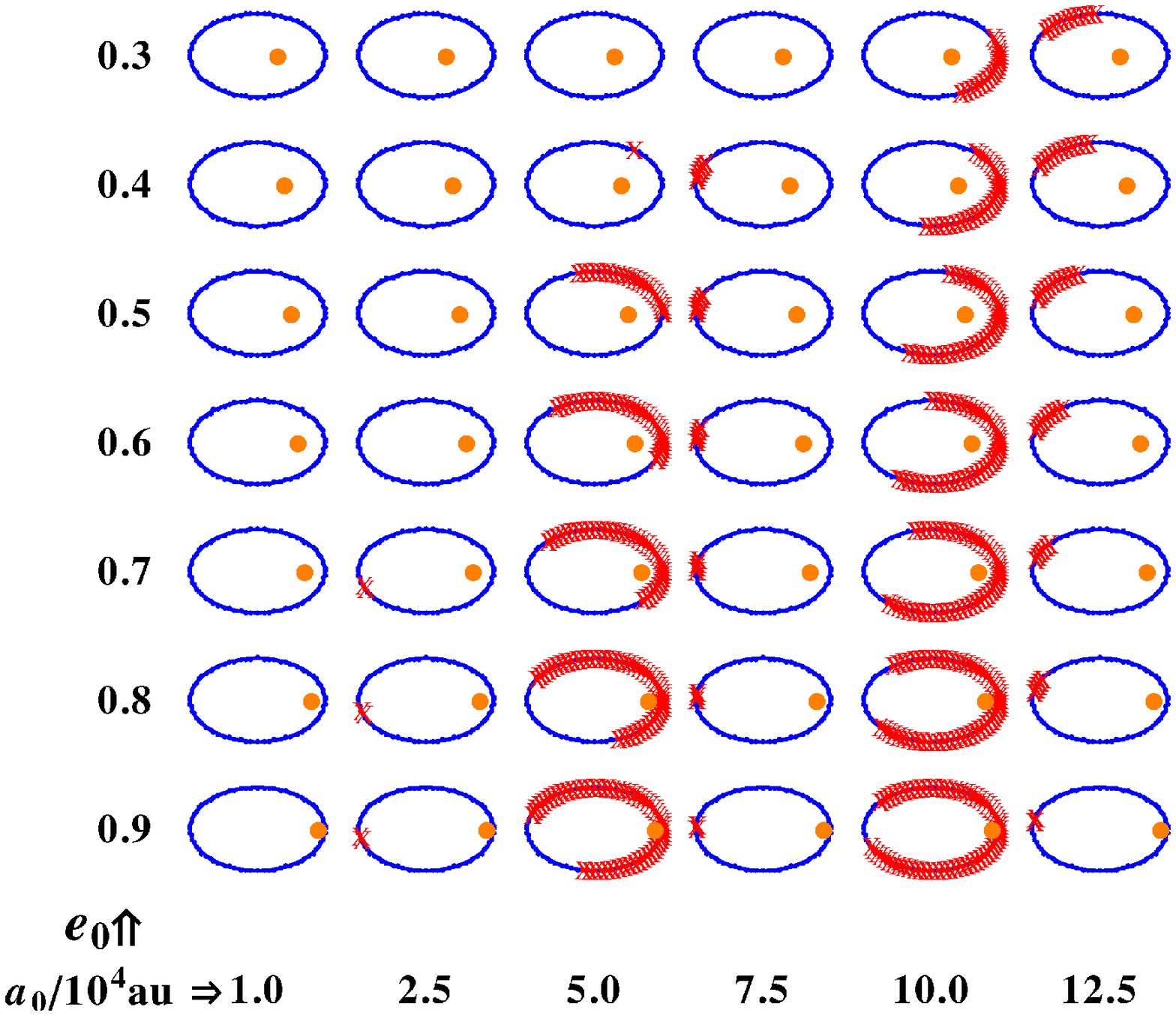,width=8.5cm,height=5.7cm}
}
\centerline{$\boldsymbol{\eta_{\odot} = 0.8}$}
\caption{
Initial condition stability snapshot for $\eta_{\odot} = 0.2$
(upper panel), $\eta_{\odot} = 0.5$ (middle panel) and
$\eta_{\odot} = 0.8$ (lower panel).   
Each scaled ellipse is for a given triplet ($\eta_{\odot},a_0,e_0$)
and graphically represent the initial $f_0$ of
unstable systems (large red crosses).  The orange dot
is a representation of the Sun, and the brown diamonds 
(at $\eta_{\odot} = 0.2$ and the highest values of $a_0$ and $e_0$) represent the values
$f_{\rm crit} = 180^{\circ} - (1/2)\cos^{-1}{(7/9)} \approx 160^{\circ}$
and $360^{\circ} - f_{\rm crit} \approx 200^{\circ}$, which, for runaway mass loss, 
bound the region in which the eccentricity must initially decrease.
For $\eta_{\odot} = 0.2$, strong ``runaway'' mass loss causes objects 
which are closest to pericenter
to be lost more easily; the effect is enhanced for higher $a_0$ and $e_0$.
For $\eta_{\odot} = 0.5$, the transitional regime between adiabatic and runaway dominates,
and objects appear to go unstable at a variety of nonconsecutive 
values of $f_0$.
For $\eta_{\odot} = 0.8$, the instability appears to be highly sensitive to $a_0$.
}
\label{plot13}
\end{figure}

\begin{figure}
\centerline{
\psfig{figure=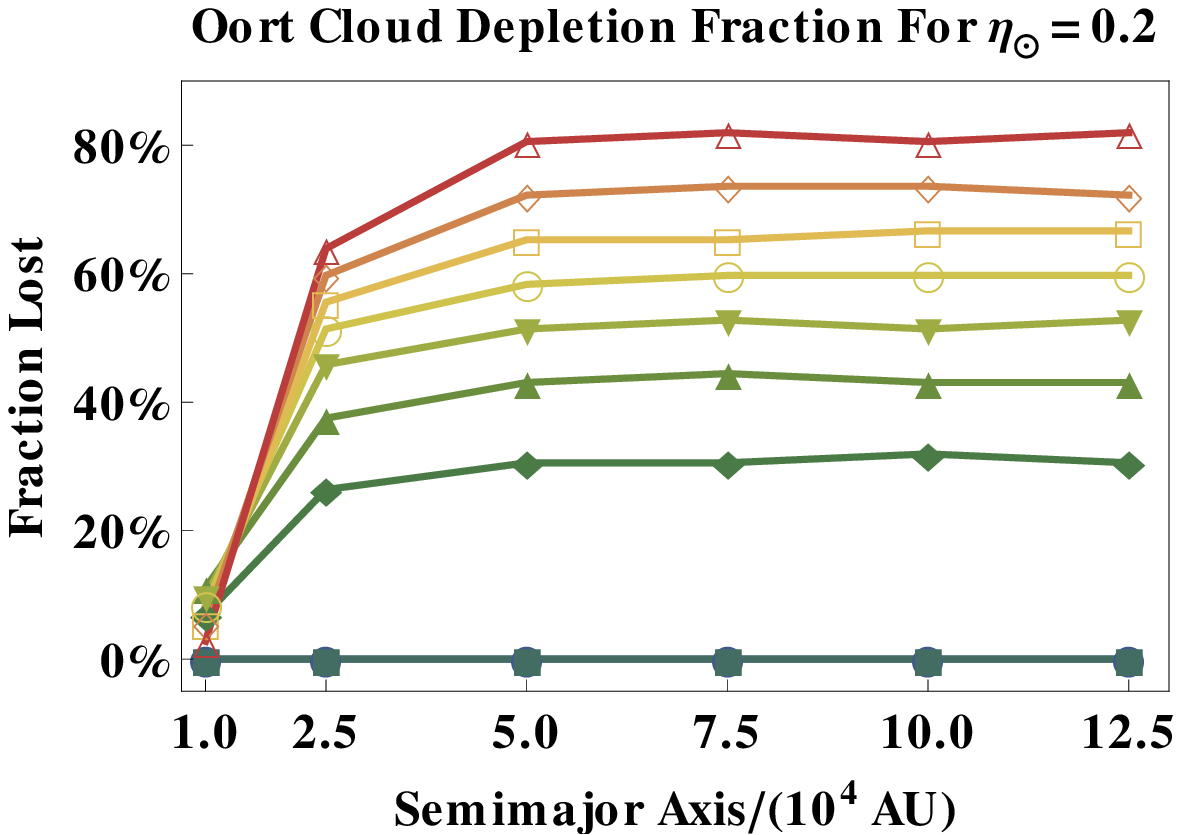,width=8.5cm}
}
\centerline{}
\centerline{
\psfig{figure=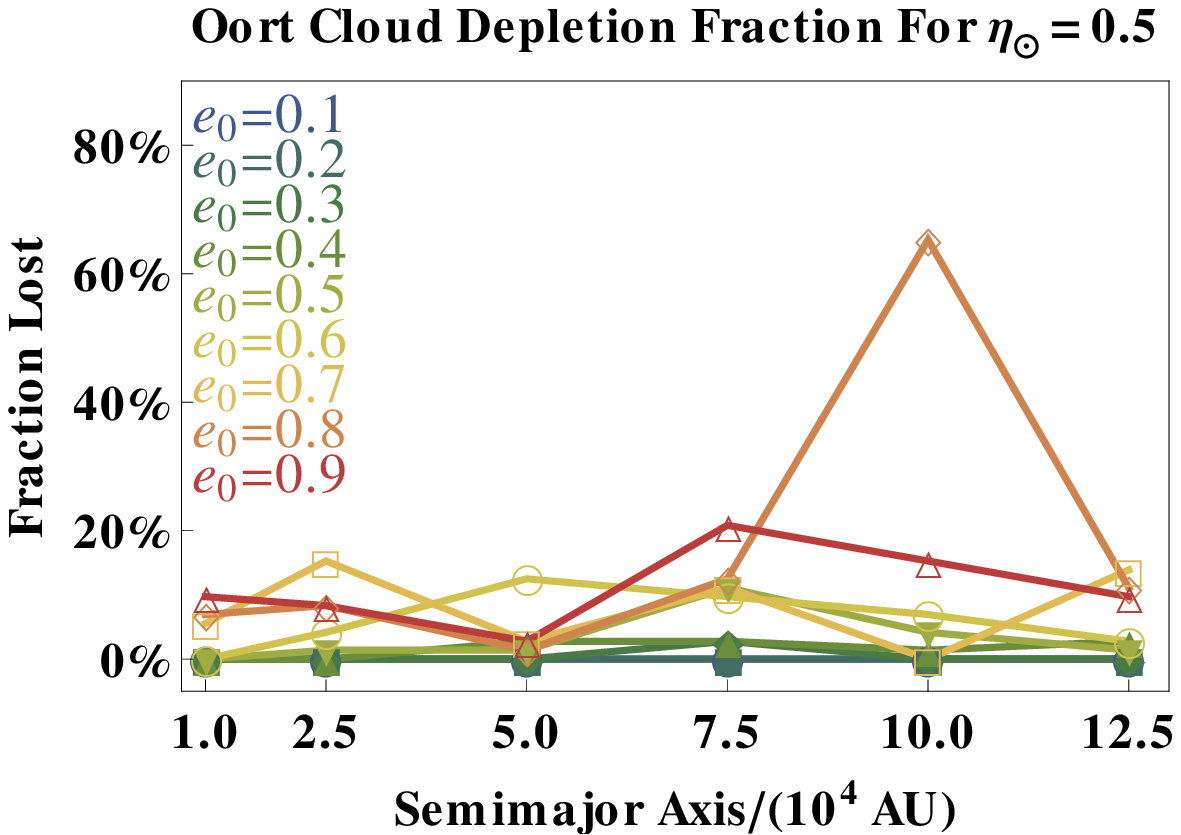,width=8.5cm}
}
\centerline{}
\centerline{
\psfig{figure=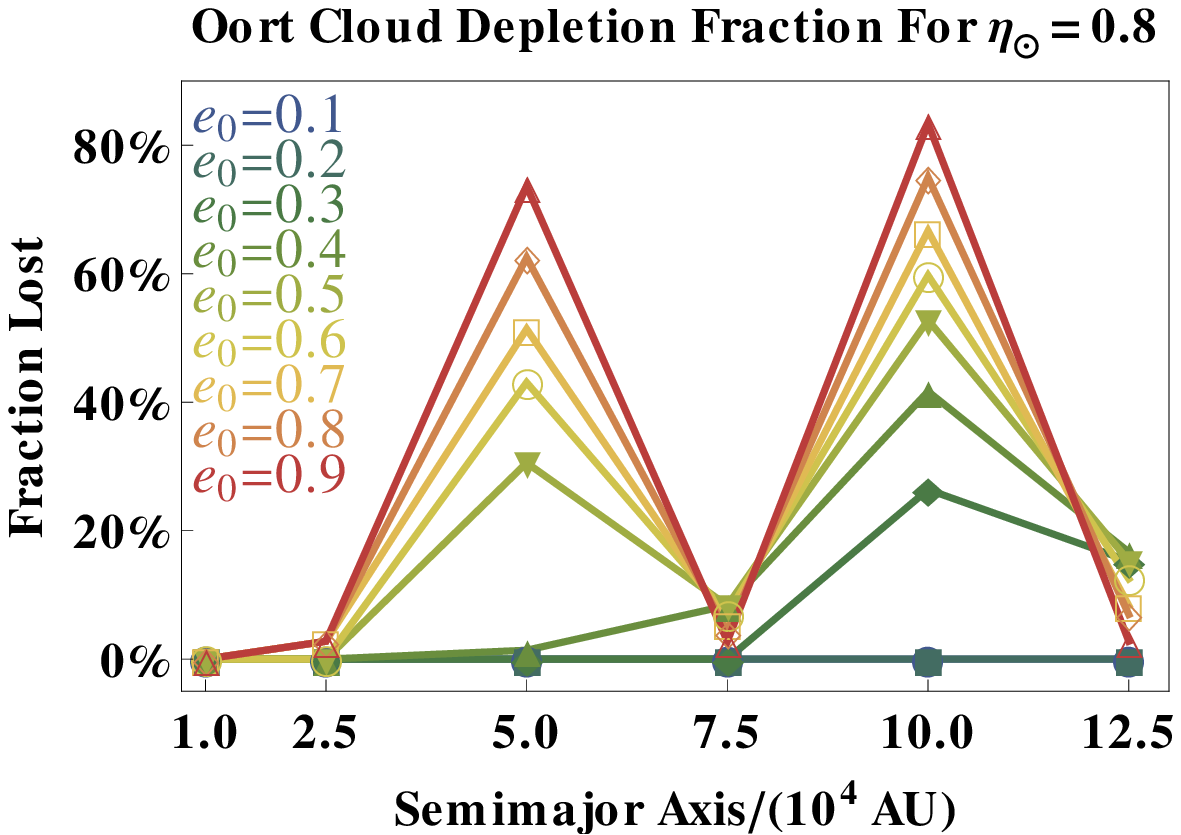,width=8.5cm}
}
\caption{Fraction of objects which escape from the Sun
for $\eta_{\odot} = 0.2$ (top panel), $0.5$ (middle panel), 
and $0.8$ (bottom panel) for the simulations described
in Section 4.2 and shown in Fig. \ref{plot13}.  
The $e = 0.9,0.8,0.7,0.6,0.5,0.4$, and $0.3$
curves are given by the colored curves with open triangles,
open diamonds, open squares, open circles, downward filled triangles,
upward filled triangles, and filled diamonds, respectively.  
Other curves do not deviate from $0\%$.  Note the stark difference in
all three distributions, demonstrating the sensitivity
of Oort Cloud depletion to both the Solar evolutionary model
and Oort Cloud model adopted.
}
\label{plot14}
\end{figure}

The results of these simulations are presented in Figs. \ref{plot13}-\ref{plot14}:
Fig. \ref{plot13} is a cartoon which illustrates the relation between instability and initial
conditions, and Fig. \ref{plot14} plots the fraction of the Oort Cloud lost depending
on the Solar evolutionary model adopted.  The cartoon indicates unstable systems with 
red crosses, as a function of $a_0$, $e_0$ and $f_0$.  The connection 
of these variables with instability helps describe the character
of the mass loss.  For example, the initial conditions which lead to instability for
the $\eta_{\odot} = 0.2$ simulations (upper panel of Fig. \ref{plot13}) are patterned and focused towards the pericenter for higher
$a_0$ and $e_0$ values, mimicking runaway regime behaviour and implying large mass loss rates (see Fig. \ref{plot6}).
In contrast, the initial conditions which lead to instability for
the $\eta_{\odot} = 0.5$ simulations (middle panel of Fig. \ref{plot13}) exhibit little order, implying that the mass
loss rates are not large enough to reside fully in the runaway regime, but are large enough
to allow for escape.  The $\eta_{\odot} = 0.8$ simulations (lower panel of Fig. \ref{plot13}) provide a third alternative, showing
a greater sensitivity to $a_0$ than to either $e_0$ or $f_0$.  The bottom two panels indicate
the difficulty in predicting the prospects for escape amidst nonlinear mass loss without appealing
to N-body simulations.

There are other features of Fig. \ref{plot13} worth noting.
Not shown are objects with $e_0 \le 0.2$, because all remain stable.
This implies that
in no case does the Sun lose enough mass for a long enough period of time
to eject distant circular bodies.  In the upper panel, at least
one object for every other $(a_0,e_0)$ pair becomes unstable.  Also, in this panel,
asymmetric signatures about the pericenter indicate that of two bodies at equal
angular separations from pericenter, the body approaching pericenter is more likely to escape.  
This asymmetry does not arise in
the analytic linear theory \citep{veretal2011}, but may be guessed based on physical grounds,
as a planet initially approaching pericenter is more likely to be ejected than a planet
initially approaching apocenter.  This tendency is also suggested in Fig. \ref{plot10} with
the top two curves in each panel.  In the bottom-rightmost ellipse in the upper panel of Fig. \ref{plot13},
we superimposed two brown diamonds at the locations of $f_{\rm crit}$ and $360^{\circ} - f_{\rm crit}$
in order to determine how closely the escape boundaries mimic the analytic boundaries which 
ensure that a body's eccentricity must initially decrease.  

The middle and bottom panels of \ref{plot13} indicate
that the fraction of bodies which escape does not necessarily scale with $a_0$; the bottom panel
displays a dramatic double-peaked red cross-dominated structure at $a_0 = 5 \times 10^4$ AU and $a_0 = 1 \times 10^5$ AU.
One can glean understanding of those panels by noting three points:  First, the orbital periods of objects 
in both panels is roughly equal to 
the mass loss timescales achieved for those Solar evolutionary tracks, placing the objects in
a transitional regime.  Second, the rate of change of eccentricity, which is proportional to $(e + \cos{f})$
\cite[see][]{veretal2011}, can assume either positive or negative values {\it and} increase {\it or} decrease
depending on the value of $a_0$ in this transitional regime.  This result is not exclusive to nonlinear
mass loss prescriptions; one can show that even for a linear mass loss rate, the inflection point of the
eccentricity with respect to time contains multiple extrema in the transitional regime.  Third, the moment
when the Sun becomes a white dwarf and stops losing mass freezes an object's orbit, even if the orbit
was approaching $e \rightarrow 1$.  Therefore, if the $\eta_{\odot} = 0.8$ post main-sequence mass loss lasted
longer, the distribution of unstable systems in the bottom panel would no longer be bimodal.

Fig. \ref{plot14} simply counts the unstable systems from Fig. \ref{plot13},
and both figures can be compared directly side-by-side.
In the top panel of Fig. \ref{plot14}, over $80\%$ of the highest eccentricity objects escape, and over half
of all objects with $e_0 = 0.5$ escape.  The escape fractions become nearly flat for
$a \ge 5 \times 10^4$ AU.  In contrast, in the middle panel, all $e_0$ curves except one
maintain escape fractions which are $< 20\%$ for all values of $a_0$ that were sampled.  
In the bottom panel, the escape fractions spike at both $a_0 = 5 \times 10^4$ AU and 
$a_0 = 1 \times 10^5$ AU for the five highest eccentricity curves. The results indicate how sensitive 
Oort Cloud depletion is to both the Oort Cloud model adopted and the Solar evolution
model adopted.

\section{Discussion}

\subsection{Planetary Perturbations}

Perturbations from surviving planets which reside within a few tens of AU of the Sun will have no effect on the orbits described here, unless an external force is evoked, or one or more of the planets is ejected from the system and achieves a close encounter on its way out.  As shown in \cite{veretal2011}, the pericenter of an orbiting isolated body experiencing isotropic mass loss {\it cannot decrease}.  This holds true in both the adiabatic and runaway regimes, for any mass loss prescription.  Therefore, mass loss cannot lower the pericenter of an Oort Cloud comet which was originally beyond Neptune's orbit.  

How the known surviving planets evolve themselves under their mutual gravitational attraction amidst post-main sequence mass loss is less clear.  With no mass loss, the giant planets are stable until the Sun turns off of the main sequence \citep{laskar2008}; \cite{khokuz2011} indicate that Jupiter and Saturn need to be about $20$ times more massive than their current values in order to suffer close encounters.  However, \cite{dunlis1998} demonstrate that with post main-sequence mass loss, giant planet evolution may not be quiescent.  In one linear mass-loss approximation, they find that Uranus and Neptune's orbits will eventually cross.  More generally, if the planets remain in orbits under $\sim 100$ AU and beyond the Sun's expanding envelope, they will evolve in the adiabatic regime, expanding their near-circular orbits all at the same rate.  They will remain bound according to Figure \ref{plot3}.  However, as multiple planets expand their orbits and maintain their relative separations, the critical separations at which the planets will remain stable will vary \citep{debsig2002}.  Also, as alluded to in Section 1, Jupiter and Saturn are predicted to largely maintain their current orbits until the Sun turns off of the main sequence.  If this is true, then they will still be close to the $5$:$2$ mean motion commensurability when significant stellar mass loss commences.  How near-resonance behavior is linked to a potentially varying critical separation is not yet clear, and suggests that much is yet to be discovered about post main-sequence planetary systems with multiple planets.

\subsection{Beyond the Solar White Dwarf}

After the Sun has become a white dwarf, surviving object orbits will retain the orbital parameters achieved at the moment the Sun stopped losing mass.  These objects will forever remain on these orbits unless subjected to additional perturbations.  For the surviving planets, the primary perturbations will come from one another, which could lead to instabilities on Gyr, or longer, timescales \citep[see, e.g., Fig. 29 of][]{chaetal2008}.  For objects further afield, the dominant perturbations will be external, and come from, for example, Galactic tides \citep[e.g.,][]{tremaine1993} \footnote{which scales as $M_{\star}(t)^{-2/3}$ \citep{tremaine1993}  and might harbor an entirely different functional form from the present-day prescription due to the collision of the Milky Way and Andromeda before the Sun turns off of the main sequence \citep{coxloe2008}.} and passing stars \citep[e.g.][]{veretal2009}.  These perturbative sources will likely play a larger role post main-sequence than during the main sequence because of orbit expansion due to mass loss.  These effects may eventually unbind the object.

Define $a_{\rm ext}$ to be the minimum semimajor axis at which an external perturbation will eventually (at some arbitrary future time) remove an orbiting object.  The brown dashed curve in Fig. \ref{plot4} indicates that the Sun will lose $\approx 48\%$ of its current mass.  Therefore, an object currently with a semimajor axis of  $a_0 \ge 0.52 a_{\rm ext}$ evolving adiabatically during post-main sequence Solar mass loss will be eventually be subject to escape from these perturbations.  However, if the object is in the runaway or transitional regime, then $a_0$ will increase by a greater amount.  For example, the red body in the left panel of Fig. \ref{plot9} increases its semimajor axis by two orders of magnitude.  In the idealized case of linear mass loss for a purely runaway object on a circular orbit at apocenter, Eq. (44) of \cite{veretal2011} gives $a_0 = 0.08 a_{\rm ext}$ assuming that the Sun loses $48\%$ of its mass entirely non-adiabatically in the runaway regime.  

\subsection{Exoplanet Analogues}

Although stellar evolutionary tracks are highly dependent on the zero-age main sequence mass and metallicity of a star, generally the maximum value of $\alpha$ increases as the stellar mass increases.  Therefore, for stars of approximately Solar metallicity which are more massive than the Sun, $a_{\rm crit}$ is expected to be lower than for the Solar System, and orbiting exobodies would be more prone to escape.  The methodology in this work can be applied to any extrasolar system for which a stellar evolutionary track can be estimated.   Given recent detections of objects which may be orbiting parent stars at $a > 10^3$ AU \citep{bejetal2008,legetal2008,lafetal2011}, this type of analysis may become increasingly relevant.

\section{Conclusion}

The Solar System's critical semimajor axis within which bodies are guaranteed to remain bound to the dying Sun during isotropic post-main sequence mass loss and in the absence of additional perturbations is $\approx 10^3$ AU - $10^4$ AU.  A more precise value can be obtained for a given Solar evolution model; this range encompasses many realistic evolutionary tracks.  The most important Solar evolutionary phase for dynamical ejection may be the RGB ($\eta_{\odot} \ge 0.7 $), the thermally pulsing AGB ($\eta_{\odot} \le 0.5$) or the early AGB ($0.5 \le \eta_{\odot} < 0.7$).  Objects with $a \gg a_{\rm crit}$, such as Oort Cloud comets, and those with $a \gtrsim a_{\rm crit}$, such as some scattered disc objects, may escape depending on their orbital architectures and position along their orbits.  Quantifying the fraction of the population of these objects which escape would require detailed modeling of their (unknown) orbital parameters at the start of the Sun's post-main sequence lifetime.

\section*{Acknowledgments}

We thank Klaus-Peter Schr{\"o}der for a valuable review of this work.


\label{lastpage}

\end{document}